\documentclass{ieeeaccess}
\usepackage{cite}
\usepackage{amsmath,amssymb,amsfonts}
\usepackage{algorithmic}
\usepackage{graphicx}
\usepackage{textcomp}
\usepackage{xcolor}
\usepackage{graphicx}
\usepackage{makecell}
\usepackage{soul}
\def\BibTeX{{\rm B\kern-.05em{\sc i\kern-.025em b}\kern-.08em
    T\kern-.1667em\lower.7ex\hbox{E}\kern-.125emX}}
\begin{document}

\title{Neural Rendering and Its Hardware Acceleration: A Review}
\author{\uppercase{Xinkai Yan}\authorrefmark{1,3}, 
\uppercase{Jieting Xu}\authorrefmark{1},\uppercase{Yuchi Huo}\authorrefmark{1,2}, and Hujun Bao\authorrefmark{1,2}
}

\address[1]{The State Key Laboratory of CAD\&CG,Zhejiang University,Hangzhou 310058,China}
\address[2]{Zhejiang Lab,Hangzhou 311121,China}
\address[3]{Jiangsu Key Laboratory of ASIC Design (Wuxi),Wuxi 214153,China}

\markboth
{X Yan \headeretal: Neural Rendering and Its Hardware Acceleration: A Review}
{X Yan \headeretal: Neural Rendering and Its Hardware Acceleration: A Review}


\begin{abstract}
Neural rendering is a new image and video generation method based on deep learning. It combines the deep learning model with the physical knowledge of computer graphics, to obtain a controllable and realistic scene model, and realize the control of scene attributes such as lighting, camera parameters, posture and so on. On the one hand, neural rendering can not only make full use of the advantages of deep learning to accelerate the traditional forward rendering process, but also provide new solutions for specific tasks such as inverse rendering and 3D reconstruction. On the other hand, the design of innovative hardware structures that adapt to the neural rendering pipeline breaks through the parallel computing and power consumption bottleneck of existing graphics processors, which is expected to provide important support for future key areas such as virtual and augmented reality, film and television creation and digital entertainment, artificial intelligence and metaverse. In this paper, we review the technical connotation, main challenges, and research progress of neural rendering. On this basis, we analyze the common requirements of neural rendering pipeline for hardware acceleration and the characteristics of the current hardware acceleration architecture, and then discuss the design challenges of neural rendering processor architecture. Finally, the future development trend of neural rendering processor architecture is prospected.
\end{abstract}

\begin{keywords}
Neural rendering, hardware acceleration, NRPU, MLP, ray marching, hash table.

\end{keywords}

\titlepgskip=-21pt

\maketitle

\section{Introduction}
\label{sec:introduction}
\PARstart{R}{endering} refers to the process in computer graphics of generating a 2D image from a 3D scene. In contrast, inverse rendering is the reverse process, involving the reconstruction of a 3D scene from a series of input 2D images. Traditional rendering and inverse rendering have mature software stacks and are supported by graphics processing units (GPUs), making them applicable in various scenarios such as gaming, film production, visual exhibitions, as well as in the creation and generation of 3D content. However, traditional rendering and inverse rendering still suffer from issues such as reliance on manual operation by professionals and complexity in specialized software, making it challenging to meet the demands of professionals in other industries for the generation and visualization of 3D content.

With the rapid rise and development of deep neural networks, significant progress has been made in various fields of artificial intelligence, including computer vision, image processing, and natural language processing. Consequently, the integration of deep learning with computer graphics has naturally become a hot research direction. On one hand, leveraging data-driven automated design methods based on deep neural networks can replace manual design by professionals, while also utilizing the prior features extracted from a large amount of training data to accelerate the rendering process. On the other hand, methods based on deep learning can integrate the entire inverse rendering process into gradient-based optimization, thereby achieving end-to-end training of deep neural networks. This integration results in faster inverse rendering and more robust outcomes. 

Neural rendering, as a fusion approach between deep learning and computer graphics, has emerged as a pivotal concept. Eslami \emph{et al.} \cite{b1} first introduced the notion of neural rendering by incorporating a Generative Query Network (GQN). This team employed GQN, which takes varying numbers of images and their corresponding camera parameters as input. Subsequently, it encodes complete scene information into a vector, utilizing this vector as input for the generative network to produce correctly occluded views. GQN innovatively learned a potent neural renderer from data without the need for human-annotated data. This renderer is capable of generating accurate, novel perspective scene images, although its rendering capabilities are limited. Nevertheless, it has inspired a plethora of subsequent work.Tewari \emph{et al.} \cite{b2} provided a definition for neural rendering:
Neural rendering is a method for generating images and videos based on deep learning. It combines deep neural network models with the physical knowledge of computer graphics to obtain controllable and realistic scene models, enabling control over scene attributes such as lighting, camera parameters, and pose. Compared to other areas of deep learning, the focal point of neural rendering is the fusion of physical/mathematical knowledge from traditional rendering with the design of neural network structures. Furthermore, it integrates knowledge from the field of computer graphics into the design of neural networks. Generally, neural rendering views neural networks as a universal function approximation tool. It achieves this by training on real-world scene data and constructing loss functions to simulate the approximation of physical/mathematical laws. Therefore, unlike traditional rendering, the quality of the neural network’s function approximation directly impacts the rendering quality. This implies that the quantity, distribution, and quality of the training input data are closely related to the final rendering effect. 

\section{Applications of Rendering}

Rendering refers to the process of generating an image or sequence of images from a 2D or 3D model (or scene) using computer software. In the context of computer graphics, rendering transforms the data in a virtual environment into a visual representation that can be displayed on a screen or printed. This process involves calculations for lighting, shading, texture mapping, and other visual elements to create a realistic or stylized final image. Recently, much progress has been seen in vanilla rendering and rendering plus AI.

Vanilla rendering technology focused on exploring efficient representation of 3D assets and the calculation of light transportation among 3D worlds. These works include Wang et al.~\cite{wang2022variational} explores directed bounding box reconstruction for solid mesh modeling, editing and rendering scattering effects for translucent objects~\cite{wang2022real} or waters~\cite{liu2023conical}, interactive model generation system~\cite{he2022interactive,yang2022rule}, etc.

Deep learning technology enhances parts of the rendering process for higher quality and performance. Wang et al.~\cite{wangbiophysically} propose a biologically-based skin model for heterogeneous volume rendering, Zheng et al.~\cite{zheng2023nelt} present neural light transform functions for global rendering pipeline on neural networks, enhance image resolution via neural super-resolution and frame prediction technologies~\cite{zhong2023neural,zhong2023fusesr,wu2023adaptive}, etc.
 
Deep learning-based Monte Carlo noise reduction by training a neural network denoiser through offline learning, it can filter noisy Monte Carlo rendering results into high-quality smooth output, greatly improving physics-based Availability of rendering techniques \cite{huo2021survey}, common research includes predicting a filtering kernel based on g-buffer \cite{bako2017kernel}, using GAN to generate more realistic filtering results \cite{xu2019adversarial}, and analyzing path space features Perform manifold contrastive learning to enhance the rendering effect of reflections \cite{cho2021weakly}, use weight sharing to quickly predict the rendering kernel to speed up reconstruction \cite{fan2021real}, filter and reconstruct high-dimensional incident radiation fields for unbiased reconstruction rendering guide \cite{huo2020adaptive,huo2022extension}, etc.
    
The many-light rendering framework is an important rendering framework outside the path-tracing algorithm. Its basic idea is to simplify the simulation of the complete light path illumination transmission after multiple refraction and reflection to calculate the direct illumination from many virtual light sources and provide a unified mathematical framework to speed up this operation \cite{dachsbacher2014scalable}, including how to efficiently process virtual point lights and geometric data in external memory \cite{wang2013gpu,wangimplementation}, how to efficiently integrate virtual point lights using sparse matrices and compressed sensing \cite{huo2015matrix,huo2022sparse}, and how to handle virtual line light data in translucent media \cite{huo2016adaptive}, sample important virtual point lights~\cite{jin2022virtual}, use spherical Gaussian virtual point lights to approximate indirect reflections on glossy surfaces \cite{huo2020spherical}, and more.

Automatic optimization of rendering pipelines Apply high-quality rendering technology to real-time rendering applications by optimizing rendering pipelines. The research contents include automatic optimization based on quality and speed \cite{wang2014automatic,liang2022automatic}, automatic optimization for energy saving \cite{ wang2016real,zhang2021powernet}, LOD optimization for terrain data \cite{li2021multi}, automatic optimization and fitting of pipeline rendering signals \cite{li2020automatic}, anti-aliasing \cite{zhong2022morphological}, automatic shader simplification~\cite{li2020automatic,huo2022shadertransformer}, etc.

Some works use a physically-based process to guide the generation of data for single image reflection removal \cite{kim2020single}, propagating local image features in a hypergraph for image retrieval \cite{an2021hypergraph,an2023towards,an2023topological}, managing 3D assets in \cite{park2021meshchain, ren2022minervas,ren2022supplementary}, generating physically plausible grasp pose~\cite{li2022contact2grasp} or segmantation~\cite{yin2022contour}, automatic indoor lighter design~\cite{ren2023data}, Seal-3D~\cite{wang2023seal} exploit rendering decomposition for interactive NeRF editing, using physical knowledge for 3D human reconstruction~\cite{chen2023immfusion,zhong2022normal}, etc.

\section{Current Status Review}
\subsection{Related Reviews}
Eslami \emph{et al.} \cite{b1} discuss where and how machine learning can improve classical rendering pipelines, as well as the data requirements for training neural rendering. The paper provides a brief overview of the fundamental knowledge in the fields of physical image generation and deep generative models, demonstrating the advantages and limitations of the hybrid approach of computer graphics and machine learning. The review further discusses the categorization of neural rendering methods, including control, computer graphics modules, explicit or implicit control, multimodal synthesis, and generality. It elaborates on applications related to neural rendering based on these categories, including new view synthesis, relighting, scene manipulation, and synthesis. The authors indicate that the purpose of this review is to enhance the research interest in neural rendering, thereby contributing to the development of the next generation of neural rendering and graphics applications.

In recent years, the surge in research related to Neural Radiance Fields (NeRF) \cite{b3} and its variants has confirmed the predictions of the DeepMind team. Tewari \emph{et al.} \cite{b4} provide a comprehensive overview of different scene representations for neural rendering, briefly introducing the basic principles of classical rendering pipelines and machine learning components, with a focus on advanced aspects of combining classical rendering with learnable 3D representations. The literature cited in this review mainly focuses on NeRF and its variant optimization work in the past two years, encompassing applications ranging from free-viewpoint videos of rigid and non-rigid scenes to shape and material editing, light field and relighting, and digital human avatar generation. Additionally, Tewari \emph{et al.} \cite{b4} also discuss the societal impacts brought about by neural rendering methods, including implications across industrial research and production, social ethics, and the environment.

In contrast to the aforementioned reviews, Wang \emph{et al.} \cite{b5} specifically concentrate on the forward applications of neural rendering, i.e., rendering pipeline-related applications that combine deep neural networks with rendering components. The paper provides a brief introduction to the theoretical foundations of physics-based rendering and deep generative networks, focusing on how neural networks can replace or enhance the work of renderers in traditional rendering pipelines and the pros and cons of this combination. It covers general and specific methods for applications such as occlusion generation, volume and subsurface rendering, multi-scene representation rendering, global illumination rendering, direct illumination rendering, human-related rendering, and rendering post-processing. The authors assert the viewpoint that traditional graphics rendering pipelines can be partially or completely replaced by deep learning-enhanced rendering.

\subsection{The scope of this article}
Unlike the related reviews on the applications of neural rendering algorithms, this paper focuses more on the integration of neural rendering in both forward and inverse rendering aspects with traditional rendering pipelines or the construction of novel neural rendering pipelines, emphasizing the common hardware acceleration requirements. To provide a clearer understanding, this paper first briefly outlines the theoretical foundations of neural rendering and then introduces representative research achievements of neural rendering in forward rendering, inverse rendering, and post-processing applications. Subsequently, this paper meticulously analyzes the common hardware acceleration requirements for computing and storage in neural rendering applications, presents recent works related to hardware-accelerated neural rendering, and discusses the design challenges of neural rendering processor architectures.
\section{Neural Rendering}
Given a high-quality description of a scene, traditional rendering methods can produce realistic images of various complex real-world phenomena. However, constructing high-quality scene models requires a significant amount of manual work. In contrast, neural rendering methods take images corresponding to specific scene conditions (such as viewpoint, lighting, layout, etc.) as data inputs. Through training, these methods construct a scene representation based on a “neural network” and render this representation under new scene attributes, thereby synthesizing new images. The “deep neural networks” in neural rendering approximate real functions through learning, enabling the resulting scene representation to be optimized for high-quality new images without being constrained by simple scene modeling approximations.
\subsection{Theoretical Foundations}
This section discusses the theoretical foundations of neural rendering work. It first introduces the physics-based rendering methods in traditional computer graphics and then outlines the approach of generative models based on deep neural networks and the representation of 3D scenes.
\subsubsection{Physics-Based Rendering}
Ray Tracing is the traditional graphics pipeline models the formation of images as a physical process in the real world: photons emitted from light sources interact with objects in the scene, resulting in a bidirectional scattering distribution function (BSDF) determined by geometric and material properties, which is then captured by the camera. This process, known as light transport, can be represented using the classic rendering equation \cite{b6}:
\begin{equation}L_{\mathrm{o}}\left(\boldsymbol{p}, \omega_{\mathrm{o}}, \lambda, t\right)=L_{\mathrm{e}}\left(\boldsymbol{p}, \omega_{\mathrm{o}}, \lambda, t\right)+L_{\mathrm{r}}\left(\boldsymbol{p}, \omega_{\mathrm{o}}, \lambda, t\right)\label{eq1}\end{equation}
Where $L_o$ represents the radiance emitted from a surface, it is a function of position $p$, light direction $\vec{o}$, wavelength $\lambda$, and time $t$. $L_e$ denotes the radiance emitted directly from the surface, while $L_r$ accounts for the interaction of incident light with surface reflectance.
\begin{equation} 
\begin{split} 
L_{\mathrm{r}}\left(\boldsymbol{p}, \omega_{o}, \lambda, t\right)=\int_{\Omega} f_{\mathrm{r}}\left(\boldsymbol{p}, \omega_{\mathrm{i}}, \omega_{\mathrm{o}}, \lambda, t\right)L_{\mathrm{i}}\left(\boldsymbol{p}, \omega_{\mathrm{i}}, \lambda, t\right)\\
\left(\omega_{\mathrm{i}} \times \boldsymbol{n}\right) \mathrm{d} \omega_{\mathrm{i}}
\end{split}
\label{eq2}\end{equation}
In \eqref{eq2}, $n$ represents the surface normal, $i$ denotes the incident light direction, $L_i$ represents the incident radiance, $f_r$ signifies the Bidirectional Scattering Distribution Function (BSDF), and $\Omega$ denotes the hemisphere surrounding the surface point. The classical rendering equation is an integral equation, and the most accurate approximation is based on Monte Carlo algorithms, simulating light path traversal through the scene via sampling.

Forward rendering is the process of transforming a scene, including the camera, lighting, surface geometry, and materials, into a simulated camera image. Computer graphics provides various approximations for the rendering equation, with the two most common forward rendering methods being rasterization and ray tracing. Rasterization, a forward process, essentially converts the 3D scene into triangles, then projects them onto the screen using perspective projection, thereby transforming the 3D representation of triangles into a 2D representation. Ray tracing, on the other hand, is a recursive process where rays are cast from image pixels into a virtual scene, simulating reflection and refraction by recursively casting new rays from intersection points. Currently, mainstream GPUs primarily accelerate the rendering process based on rasterization, as it is parallel computation-friendly, suitable for hardware acceleration, and exhibits good memory coherence. However, due to the superior effects of ray tracing in global illumination and other complex lighting scenarios (such as depth of field and motion blur), hardware manufacturers have recently added ray tracing hardware acceleration structures to their GPUs, thereby supporting ray tracing in real-time rendering pipelines \cite{b7,b8}. This allows for a blend of rasterization and ray tracing to achieve better rendering effects.

Inverse rendering, the process of estimating different model parameters (such as camera, geometry, material, and light parameters) from real data to generate new views or edit materials or lighting, addresses the problem of constructing a 3D scene that generates a given image. Due to mathematical complexity or computational expense, predefined physical models or data structures used in classical rendering do not always accurately reproduce all features of real-world physical processes. This constitutes a primary challenge in inverse rendering. Deep neural networks, however, can statistically approximate the physical processes of inverse rendering, leading to output data that more closely resembles training data, thereby enabling a more accurate representation of real-world effects.
\subsubsection{Deep Neural Networks}
Traditional computer graphics methods focus on modeling scenes using physical/mathematical formulas and simulating light transport to generate images. Deep neural networks, however, address this problem from a statistical probability perspective by learning the image distribution in the real world.

In the field of machine learning, the multilayer perceptron (MLP) is commonly utilized as a universal function approximator. It is a traditional fully connected neural network that takes spatial coordinates as input within the context of scene representation and produces corresponding output values. This type of network, also known as a coordinate-based neural network, serves as a function approximator for representing surfaces or volumes, as well as storing other properties. Since NeRF, a substantial amount of neural rendering work has adopted MLP as a function approximator.

Deep generative networks excel at generating random realistic images that resemble statistical data from the training set. While traditional generative adversarial networks (GANs) synthesize virtual images from random vectors resembling the training data, this is insufficient for scene rendering. Current neural rendering employs deep generative networks using perceptual distance for training or utilizes conditional generative adversarial nets (cGANs), both of which employ pre-trained auxiliary networks to define an effective training objective, resulting in improved generative networks and rendering effects.

Originating from the Transformer model proposed in natural language processing, the self-attention mechanism reduces the distance between any two positions in a sequence to a constant, thereby addressing long-range dependency issues. Through matrix calculations, it directly computes the correlation between each word without the need for passing through hidden layers. This concept inspires and simplifies the task of neural rendering using MLP to learn the mapping from position to specific values, and it can also be used for geometric and appearance inference.
\subsubsection{Three-Dimensional Scene Representation}
To model objects in a 3D scene, various methods of scene representation have been proposed, mainly categorized into explicit and implicit representations. Both explicit and implicit methods can be used for surface and volume representations of scenes. Mainstream 3D data formats include depth, voxels \cite{b13}, points, and meshes.

A depth map contains information related to the distance from the viewpoint to objects in the scene. The channel itself is similar to a grayscale image, where each pixel value represents the actual distance from the sensor to the object. Since a depth map contains information only about the foremost object surfaces in the scene, additional visibility calculations are unnecessary. Furthermore, depth maps are very friendly to neural networks and serve as a format well-suited for training and computation through neural networks.

Voxels are commonly used to represent volumes. They can store geometric occupancy, density values of scenes with volume effects (such as transparency), and the appearance of the scene \cite{b14}. Traditional rendering seldom utilizes voxels because it requires both visibility and shadow calculations. However, voxels are relatively friendly to neural rendering, as neural networks can use voxels to learn end-to-end rendering pipelines.

A point cloud is a set of elements in Euclidean space, where the continuous surface of a scene can be discretized using a point cloud. Each element of the point cloud represents a sample point (x, y, z) on the surface. Additionally, point clouds can also represent volume (such as storing opacity or density values). In neural rendering, point clouds can store some learnable features \cite{b15,b16}, thereby projecting and fusing individual features into 2D feature maps, mainly applied in denoising, stylization, and novel view synthesis \cite{b16}--\cite{b18}. However, due to the association of point cloud data with the quantity, order, and sampling of surface points, it is relatively unfriendly to neural networks.

Polygon mesh represents a piecewise linear approximation of a surface and is widely used as a standard representation in most graphic editing tools in traditional computer graphics. To be compatible with traditional rendering pipelines, many neural rendering methods use mesh representation.
\subsubsection{Neural Scene Representation}
In the context of neural rendering, using neural networks to approximate the surface or volume representation function of a scene is referred to as neural scene representation. Both surface and volume representations can store additional information, such as color or radiance.

The signed distance function (SDF) is the most commonly used implicit surface representation. It returns the shortest distance from any point in space to the object surface. All points with a distance of 0 represent the object surface, points with a distance less than 0 represent the interior of the object, and points with a distance greater than 0 represent the exterior.

Neural implicit surfaces \cite{b19,b20,b21,zhusupplementary,zhu2022learning, xie2023holistic} in neural rendering are obtained by combining the signed distance function with Multilayer Perceptron (MLP) and are used for shape modeling. MLP maps continuous coordinates to signed distance values. This representation method has been widely applied in neural scene representation and rendering. Neural implicit surface representation has many advantages, such as high memory efficiency and theoretically being able to represent geometry at infinite resolution.
\Figure[t!](topskip=0pt, botskip=0pt, midskip=0pt)[width=1.0\linewidth]{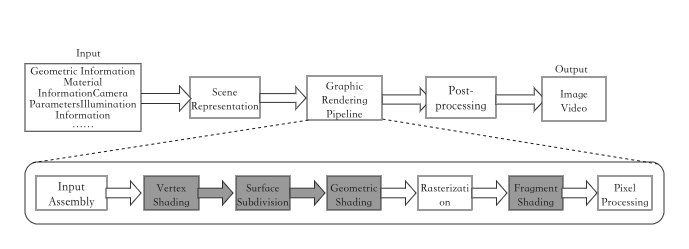}
{\textbf{Pipeline of forward rendering.}\label{fig1}}
Similar to neural implicit surfaces, neural voxels represent voxels using neural networks instead of using voxel grids to store features or other variables. In neural rendering, coordinate-based neural networks are generally used to model scenes volumetrically. Since MLP networks can be used to parameterize volumes, neural voxel representation may be more efficient than explicit voxel grid representation.

Volume rendering \cite{b13} is based on ray casting and is mainly used for rendering non-rigid objects such as clouds and smoke. It divides the process of photon-particle interaction into four types: absorption, emission, external scattering, and internal scattering. NeRF \cite{b3} and related works \cite{b22}--\cite{b27} have demonstrated good performance in learning scene representation from multi-view input data using volume rendering, which can be utilized in neural rendering-based inverse rendering frameworks.
\subsection{Application}
This section briefly introduces the latest or representative applications of neural rendering technology in the three categories of forward rendering, inverse rendering, and post-processing.
\subsubsection{Forward Rendering}
Fig. \ref{fig1} illustrates the traditional forward rendering pipeline, which consists of three stages: scene representation, graphics rendering pipeline, and post-processing. The role of neural rendering technology in forward rendering typically involves enhancing various sub-modules of the traditional rendering process, such as voxel scene representation and global illumination.

\begin{table*}[!ht]
    \centering
     \caption{The Application of Neural Rendering in Forward Rendering}
    \label{table1}
    \begin{tabular}{|l|l|l|l|l|}
    \hline
        Stage Classification & Submodule Classification & Work & Input Data Type & Generalization \\ \hline
        Scene Representation & Voxel-based & render net\cite{b28} & Voxel & Generalized \\ 
        ~ & ~ & ~ & Camera Pose & ~ \\ 
        ~ & ~ & ~ & Light Source Position & ~ \\ \cline{3-5}
        ~ & ~ & neural voxel renderer\cite{b29} & Voxel & Generalized \\ 
        ~ & ~ & ~ & Camera Pose & ~ \\ 
        ~ & ~ & ~ & Light Source Position & ~ \\ \cline{2-5}
        ~ & Point Cloud-based  & neural point-based graphic\cite{b30} & Point Cloud & Generalized \\ 
        ~ & ~ & ~ & Descriptors & ~ \\ 
        ~ & ~ & ~ & Camera Pose & ~ \\ \cline{3-5}
        ~ & ~ & neural point cloud rendering\cite{b31} & Point Cloud & Specialized \\ 
        ~ & ~ & ~ & Camera Pose & ~ \\ \cline{3-5}
        ~ & ~ & neural light transport\cite{b32} & Point Cloud & Specialized \\ 
        ~ & ~ & ~ & Scene Features & ~ \\ \cline{2-5}
        ~ & Network-based  & differentiable volumetric\cite{b33} & Pixel & Specialized \\ \cline{3-5}
        ~ & ~ & neural implicit surfaces\cite{b34} & Image Set/SDF & Specialized \\ \hline
        Global Illumination & Indirect Illumination  & neural light transport\cite{b32} & Point Cloud & Specialized \\ 
        ~ & ~ & ~ & Scene Features & ~ \\ \cline{3-5}
        ~ & ~ & deep illumination\cite{b35} & Normal Map & Generalized \\ 
        ~ & ~ & ~ & Indirect Illumination & ~ \\
        ~ & ~ & ~ & Diffuse Map & ~ \\
        ~ & ~ & ~ & Depth Map & ~ \\ \cline{3-5}
        ~ & ~ & neural control variates\cite{b36} & Camera Pose & Generalized \\
        ~ & ~ & ~ & Probability Density & ~ \\ \cline{2-5}
        ~ & Direct Illumination & neural screen space rendering\cite{b37} & Material & Generalized \\
        ~ & ~ & ~ & Geometric Shape & ~ \\
        ~ & ~ & ~ & Lighting & ~ \\ \hline
        Spatial Effects & Ambient Shadows & deep shading\cite{b38} & Position & Generalized \\ 
        ~ & ~ & ~ & Normal & ~ \\ 
        ~ & ~ & ~ & Reflectance & ~ \\ 
        ~ & ~ & ~ & Color & ~ \\ \cline{2-5}
        ~ & Ambient Shadows & AOGAN\cite{b39} & Position & Generalized \\
        ~ & ~ & ~ & Normal & ~ \\ \hline
    \end{tabular}
\end{table*}

Table. \MakeUppercase{\romannumeral 1} presents relevant research on neural rendering in various rendering sub-modules and the input data types for the networks.
\Figure[t!](topskip=0pt, botskip=0pt, midskip=0pt)[width=1\linewidth]{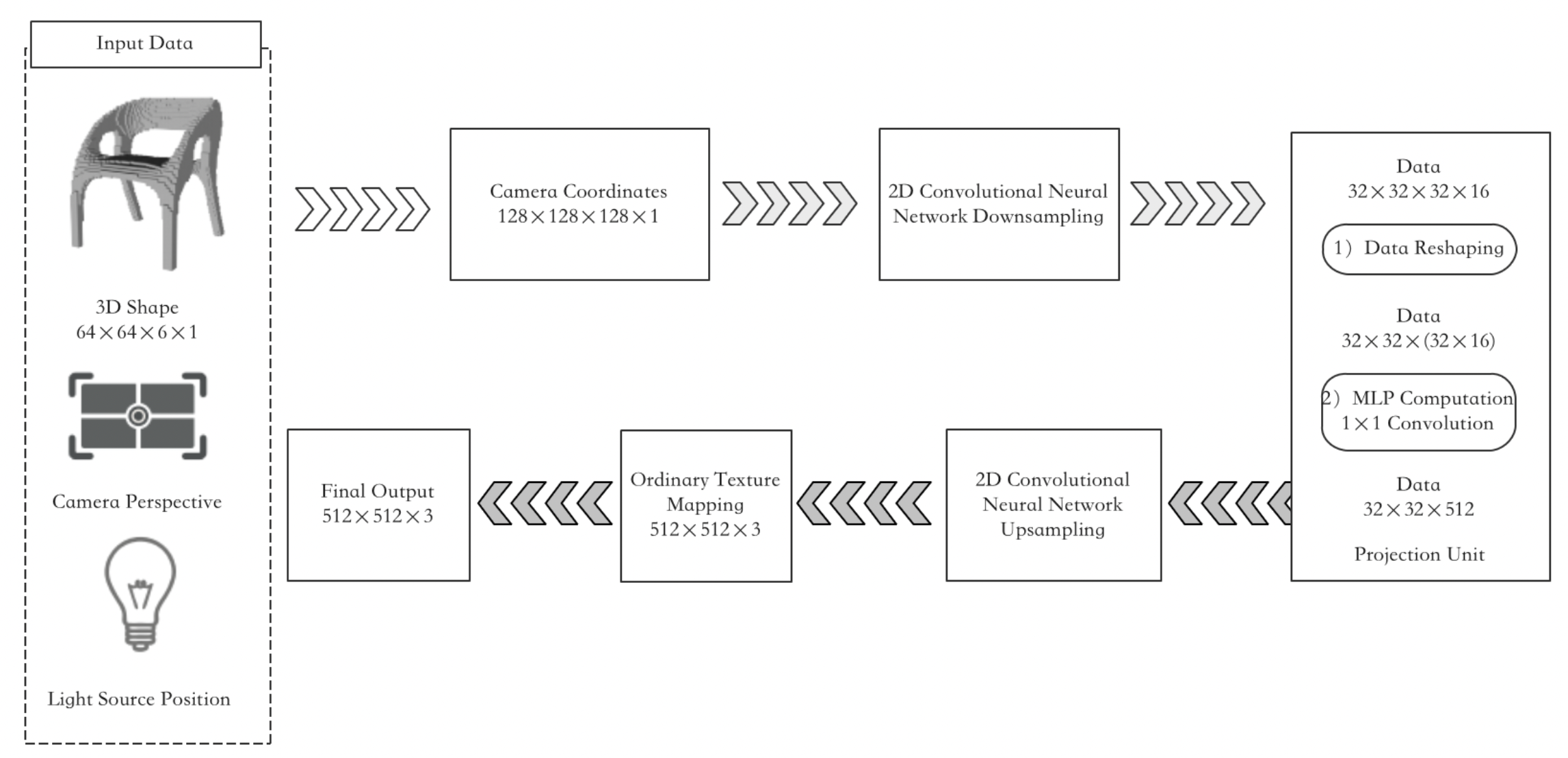}{\textbf{The network architecture of render net.}\label{fig2}}
Voxel-based Scene Representation. Fig. \ref{fig2} depicts the end-to-end neural voxel rendering process implemented by Render Net \cite{b28}. Initially, it transforms the input voxel grid, camera pose, and light source position into camera coordinates. Subsequently, a 3D convolutional neural network (3D CNN) converts the camera coordinates into a 4D tensor (H, W, D, C) representing neural voxels. Following this, a projection unit transforms the neural voxels into a 3D tensor representing neural pixels (with matrix dimensionality transformation for the depth dimension D and feature channel dimensionality C). Finally, a 2D CNN network performs deconvolution operations to render the projected neural voxels into images. Neural voxel renderer \cite{b29} proposes a deep learning-based rendering method to map voxelized scenes to high-quality images. The authors design two neural renderers, NVR and NVR+, for rendering scenes. NVR and Render Net are quite similar, with the distinction that NVR’s projection unit uses a 2-layer MLP to process light positions, thereby encoding lighting information. NVR+ is discussed in Section \ref{1}.
\Figure[t!](topskip=0pt, botskip=0pt, midskip=0pt)[width=1\linewidth]{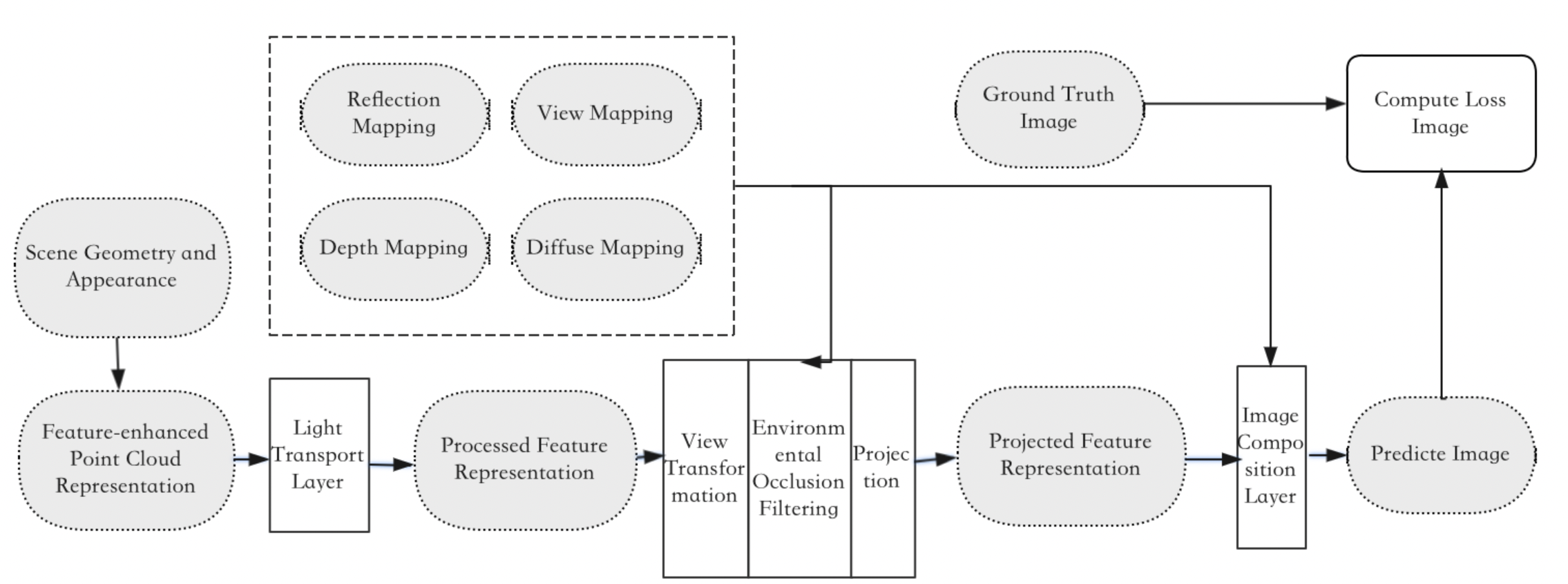}
{\textbf{The model architecture of neural light transport.}\label{fig3}}
Point Cloud-based Scene Representation. Neural point-based graphics \cite{b30} introduces a method for modeling the appearance of real scenes based on point clouds. The authors use the original point cloud as the geometric representation of the scene and enhance each point cloud through encoding local geometry and appearance, using pre-trained neural descriptors. The article uses Z-Buffer to rasterize points at several resolutions as input data for the neural rendering network, employing neural descriptors as pseudo-colors. The network architecture is similar to U-Net, rendering images through rasterization. This model adapts to new scenes by optimizing the parameters of the neural rendering network and propagating the loss function through backpropagation of the neural descriptors. Neural point cloud \cite{b31} further proposes a neural point cloud rendering pipeline using multi-plane projection (MPP). The authors facilitate the automatic learning of the visibility of 3D points by projecting 3D points into a layered volume within the camera frustum. The network framework consists of two modules: voxelization based on multiple planes and multi-plane rendering. The voxelization module uniformly divides the 3D space of the camera frustum into small voxels based on the image size and a predefined number of planes, aggregating each small frustum and generating a multi-plane 3D representation. The plane rendering module, as the rendering network, predicts a 4-channel output (RGB + blending weight) for each plane, ultimately blending all planes based on the blending weights to produce the rendered image. Neural light transport \cite{b32} proposes a method using neural networks to learn light transport in static and dynamic 3D scenes. Fig. \ref{fig3} illustrates the specific model structure of NLT. The neural rendering network architecture primarily comprises three modules: light transport layer, transformation and projection layer, and image synthesis layer. The light transport layer uses MLP to achieve the transformation from feature-rich point clouds to processed features, while the image synthesis layer employs a U-Net-like architecture with residual blocks to achieve image rendering. In comparison with existing 2D image-domain methods, the proposed approach supports inference in both 3D and 2D spaces, thereby achieving global illumination effects and 3D scene geometry processing. Additionally, this model can produce realistic rendering of static and dynamic scenes.

Based on network-based scene representation. Differentiable volumetric rendering (DVR) \cite{b33} proposes a differentiable rendering formula for implicit shape and texture representation. The authors devise an occupancy network that assigns occupancy probabilities to each point in 3D space, using isosurface extraction techniques to extract object surfaces, while directly learning implicit shape and texture representations from RGB images using texture fields \cite{b40}. Yariv \emph{et al.} \cite{b34} define the volume density function as the cumulative distribution function (CDF) of the Laplacian, applied to Signed Distance Function (SDF) representations. The authors argue that this simple density representation has three advantages: first, it provides useful inductive biases for learning geometry in the neural volume rendering process; second, it aids in constraining approximation errors in opacity, leading to precise sampling of observed light rays; and third, it allows effective unsupervised disentanglement of shape and appearance in voxel rendering. The network framework consists of two Multilayer Perceptrons (MLPs): the first MLP is used to approximate geometric SDF and 256-dimensional global geometric features, while the second MLP is used to represent the scene radiance field.
\Figure[t!](topskip=0pt, botskip=0pt, midskip=0pt)[width=1\linewidth]{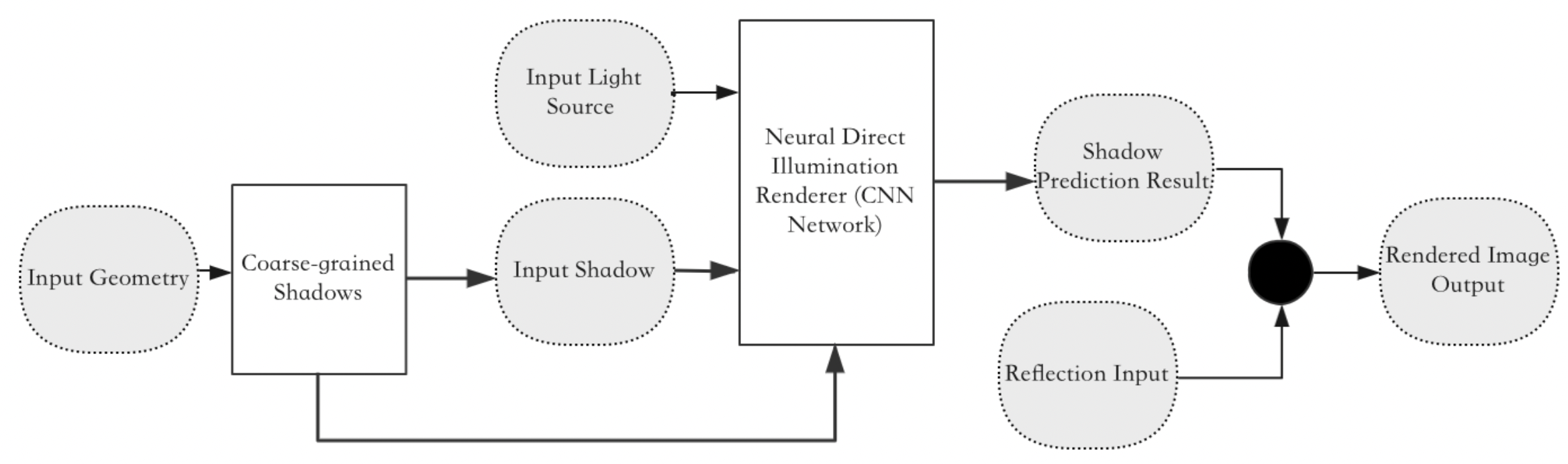}
{\textbf{The model architecture of neural direct-illumination renderer.}\label{fig4}}
Global illumination. Deep illumination (DI) \cite{b35}, in real-time applications, approximates global illumination using conditional Generative Adversarial Networks (cGAN). The authors propose a mapping network from the G-buffer (depth, normal, and diffuse albedo) and direct illumination to any global illumination, where the generative model can be used to learn the density estimation of advanced lighting models from the screen-space buffer to the 3D environment. Neural control variates (NCV) \cite{b36} propose a method to reduce unbiased variance in parameter Monte Carlo integration. The authors use a neural network to learn a function close to the rendering equation, a neural sampler to generate sampling probabilities, and a neural network to infer the solution of the integral equation. The neural direct-illumination renderer (NDR) \cite{b37} can render direct illumination images of any geometric shape with opaque materials under distant lighting. Fig. \ref{fig4} illustrates the specific model structure of NDR. First, the screen-space buffer (material, geometric shape, and lighting) is input into a CNN network to learn the mapping from pixel buffer to rendered image, then NDR predicts the shadow map and combines it with the irradiance map to render the image.

Environment Shading. Deep shading \cite{b38} introduces a novel technique for generating various rendering effects, including environment shading using deferred shading buffers and convolutional neural networks. The authors employ deep neural networks as renderers, taking deferred shading buffers as input to produce specific rendering outcomes. AOGAN \cite{b39} presents an end-to-end generative adversarial network for producing realistic environment shading. The authors explore the significance of perceptual loss in the generation model's accuracy for environment shading by combining VGG-structured perceptual loss with GAN-structured adversarial loss. Additionally, they introduce a self-attention module, taking position and normal shading buffers as input, to enhance the training's generalization.
\subsubsection{Inverse Rendering}
Fig. \ref{fig5} illustrates the schematic of the inverse rendering pipeline, divided into three stages: scene observation, inverse rendering pipeline, and post-processing. The role of neural rendering techniques in inverse rendering typically involves utilizing deep neural networks for tasks such as novel view synthesis, relighting, and material editing.
\Figure[t!](topskip=0pt, botskip=0pt, midskip=0pt)[width=1\linewidth]{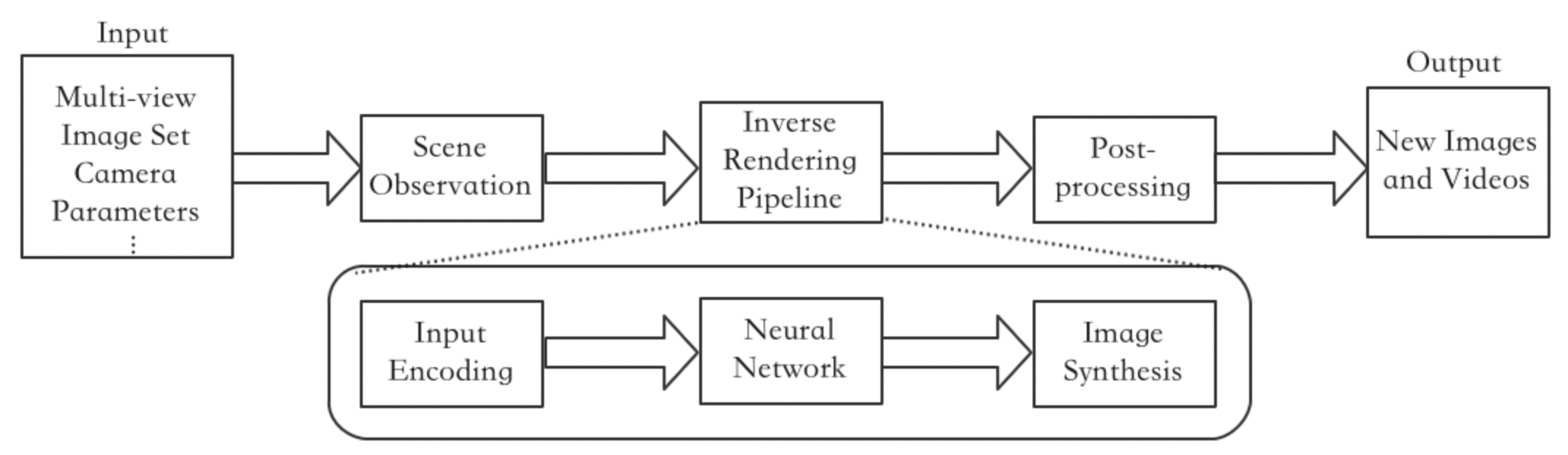}
{\textbf{The pipeline of inverse rendering.}\label{fig5}}
\begin{table*}[!ht]
    \centering
    \caption{The Application of Neural Rendering in Inverse Rendering}
    \label{table2}
    \begin{tabular}{|l|l|l|l|l|}
    \hline
        Functionality Classification & Work & Generation Model & 3D Scene Representation & Input Data Type \\ \hline
        New View Synthesis & neural voxel renderer\cite{b29} & U-Net & Voxel & Voxel \\ \cline{2-5}
        (Voxel-based) & deep voxels\cite{b41} & U-Net & Voxel & Voxel \\ \cline{2-5}
        ~ & neural volumes\cite{b42} & None & Voxel & Image \\ \hline
        New View Synthesis & NeRF\cite{b3}, autoint\cite{b43} & None & MLP & Image \\ \cline{2-5}
        (NeRF-based) & ~ & ~ & ~ & Camera Pose \\ \cline{5-5}
        ~ & neural sparse voxel fields\cite{b44},kiloNeRF\cite{b45} & None & Neural Voxel + Grid & Image \\ \cline{2-5}
        ~ & ~ & ~ & ~ & Camera Pose \\ \cline{2-5}
        ~ & baking NeRF\cite{b46},fastNeRF\cite{b47},GeoNeRF\cite{b48}, & None & Grid & Image \\ \cline{5-5}
        ~ & plenoctrees\cite{b49},light field networks\cite{b50}, & ~ & ~ & Camera Pose \\ \cline{5-5}
        ~ & instant neural graphics primitives\cite{b51} & ~ & ~ & ~ \\ \cline{2-5}
        ~ & D-NeRF\cite{b24},neural scene flow fields\cite{b52} & None & MLP & Video \\ 
        ~ & ~ & ~ & ~ & Camera Pose \\ \cline{2-5}
        ~ & GIRAFFE\cite{b23} & CNN & Neural Feature Field & Image \\ \cline{2-5}
        ~ & pixelNeRF\cite{b53} & None & Neural Voxel & Image \\ \cline{5-5}
        ~ & ~ & ~ & ~ & Camera Pose \\ \cline{2-5}
        ~ & IBRnet\cite{b54},common objects in 3D\cite{b55} & None & Neural Voxel & Image \\ \cline{5-5}
        ~ & ~ & ~ & ~ & Camera Pose \\ \cline{1-5}
        New View Synthesis and & deepSDF\cite{b19} & CNN & SDF & Image \\ \cline{5-5}
        Scene Reconstruction & ~ & ~ & ~ & Camera Pose \\ \cline{2-5}
        (SDF-based) & I2-SDF\cite{b21} & MLP & SDF & Image \\ \cline{5-5}
        ~ & ~ & ~ & ~ & Camera Pose \\ \cline{2-5}
        ~ & scene representation networks\cite{b56} & CNN & SDF & Image \\ \cline{5-5}
        ~ & ~ & ~ & ~ & Camera Pose \\ \hline
        Relighting and & I2-SDF\cite{b21} & MLP & SDF & Image \\ \cline{5-5}
        Material Editing & ~ & ~ & ~ & Camera Pose \\ \cline{2-5}
        ~ & neural reflectance fields\cite{b57} & CNN & Neural Voxel & Image \\ \cline{5-5}
        ~ & ~ & ~ & ~ & Light Parameters \\ \cline{2-5}
        ~ & neural light transport\cite{b58} & MLP & Neural Voxel & Image \\ \cline{5-5}
        ~ & ~ & ~ & ~ & Light Parameters \\ \cline{2-5}
        ~ & NeLF\cite{b59} & MLP & Neural Feature Vector & Portrait \\ 
        ~ & ~ & ~ & ~ & Camera Pose \\ \hline
    \end{tabular}
\end{table*}
Table. \MakeUppercase{\romannumeral 2} provides an overview of relevant research in neural rendering for inverse rendering, along with corresponding types of generative networks and 3D scene representation types.

Novel View Synthesis based on 3D Voxel Representation. The neural voxel renderer NVR+ network proposed by neural voxel renderer \cite{b29} extends the NVR network by incorporating a splatting network and a neural re-rendering network. The splatting network initially synthesizes images by splitting the colored voxel centers in the target view and then passes this image, along with the feature vector generated by NVR, to a 2D convolutional encoder. The final result is processed by the neural re-rendering network U-Net to generate the output image. Deep-voxels \cite{b41} encode the visual appearance of 3D scenes, achieving end-to-end inverse rendering. It first employs a 2D U-Net network to extract 2D feature maps, then utilizes differentiable lifting layers to raise 2D features to 3D feature volumes. Subsequently, a 3D U-Net processes the feature volume after fusion, mapping the feature volume to the camera coordinate systems of two target views through differentiable reprojection layers. An occlusion network calculates the soft visibility of each voxel, and finally, a 2D U-Net rendering network generates two final output images. Neural volume \cite{b42} introduces a learning-based approach to represent dynamic objects, employing an encoder-decoder network to convert input images into 3D voxel representations while using differentiable ray-marching operations, thus enabling end-to-end training. The innovation of neural volume lies in the fact that any input image can serve as supervision through re-rendering loss, eliminating the need for explicit reconstruction or target tracking.
\Figure[t!](topskip=0pt, botskip=0pt, midskip=0pt)[width=1\linewidth]{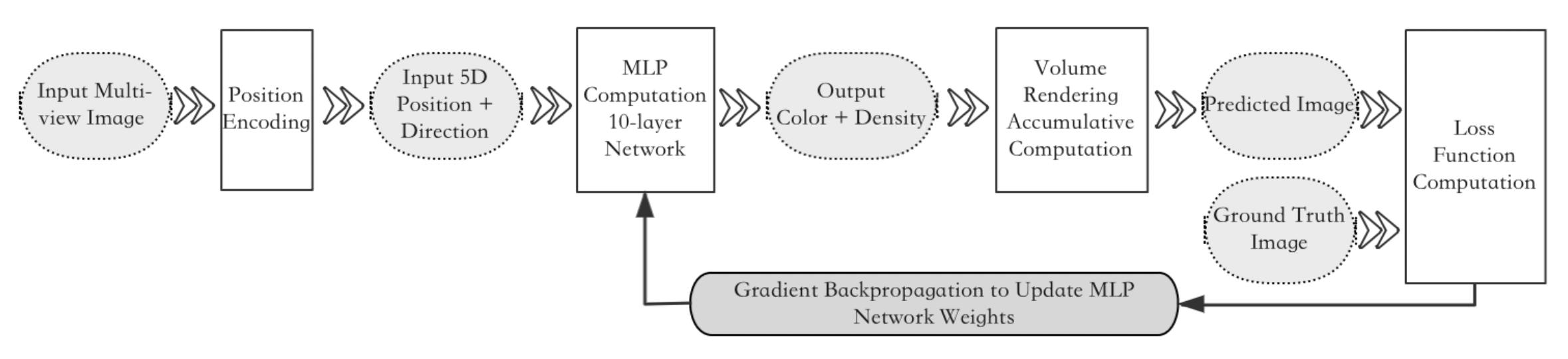}
{\textbf{The model architecture and differentiable render procedure of NeRF.}\label{fig6}}
The emergence of Neural Radiance Fields (NeRF) \cite{b3}, based on Multi-Layer Perceptron (MLP) networks for scene representation, marks a breakthrough in the application of realistic novel view synthesis within a single scene. Fig. \ref{fig6} illustrates the NeRF model structure and its differentiable rendering process. NeRF directly applies voxel rendering to synthesize images from the MLP. The MLP maps positions and directions to volume density and color. Subsequently, it optimizes the MLP weights based on pixel-level rendering losses from input images to represent each new input scene. The primary innovation of NeRF lies in the introduction of positional encoding, enabling effective differential compression of scenes during the optimization process, thereby achieving higher resolution compared to discrete 3D voxel representations without increasing the number of MLP weights. However, the computation of color and volume density for individual points within NeRF introduces significant computational overhead, resulting in slow rendering speeds. Additionally, NeRF suffers from poor generalization, supports only static scenes, and requires extensive sample training.

Several efforts \cite{b44}--\cite{b47} \cite{b49} have focused on optimizing the NeRF architecture to improve rendering speeds. AutoInt \cite{b43} and Light Field Networks \cite{b50} enhance rendering speeds by training the MLP itself. Müller \emph{et al.} \cite{b51} optimize NeRF training speed by introducing multi-resolution hash encoding to alleviate MLP training burdens, thereby accelerating training speeds.

Further enhancements based on NeRF have been explored. Pumarola \emph{et al.} \cite{b24} and Li \emph{et al.} \cite{b52} have improved NeRF for static scenes. D-NeRF \cite{b24} introduces time as an input, dividing the learning process into two main stages: encoding scenes into canonical space and mapping canonical representations to deformed scenes at specific times, utilizing fully connected networks. Meanwhile, neural scene flow fields \cite{b52} model dynamic scenes as time-varying continuous functions of appearance, geometry, and 3D scene motion. In recent years, numerous efforts have focused on optimizing the generalization of NeRF. GIRAFFE \cite{b23} learns local features for each pixel in 2D images and projects these features into 3D points, thereby generating a universal and rich point representation. GeoNeRF \cite{b48} utilizes Transformers to render and infer geometry and appearance, while using volume rendering to capture image details, thus achieving a NeRF-based generalized realistic novel view synthesis method. PixelNeRF \cite{b53} aggregates features across multiple views within a voxel rendering framework, thereby rendering in a NeRF-like manner to produce color and density fields using MLP. When trained on multiple scenes, they learn scene priors for reconstruction, enabling high-fidelity scene reconstruction from a few viewpoints. IBRNet \cite{b54} introduces Ray-Transformer networks to learn a universal view interpolation function, supporting generalization to new scenes. NeRFormer \cite{b55} employs Transformer networks to reconstruct objects with few viewpoints.

Based on SDF for new view synthesis. DeepSDF \cite{b19} proposes a learnable continuous Signed Distance Function (SDF) representation, enabling high-quality shape representation, interpolation, and completion from partial and noisy 3D input data. Scene Representation Networks (SRNs) \cite{b56} primarily consist of an MLP scene representation module, a ray-stepping LSTM module, and a pixel generator module. The authors achieve end-to-end training of a set of images without explicit supervision based on the SDF implicit function, enabling new view synthesis, few-shot reconstruction, shape and appearance interpolation. I2-SDF \cite{b21} recovers basic shape, incident radiance, and material from multi-view images based on an overall neural SDF framework. By employing surface-based differentiable Monte Carlo ray tracing and radiance primitive segmentation, the neural radiance field is decomposed into spatially varying material within the scene, demonstrating superior quality in indoor scene reconstruction, new view synthesis, and scene editing.

Relighting and material editing. Neural Reflectance Fields \cite{b57} propose the first extension of NeRF to achieve relighting, representing the scene as a volume density field, surface normals, and Bidirectional Reflectance Distribution Functions (BRDFs) field, allowing scene rendering under arbitrary lighting conditions. The method evaluates the amount of incoming light rays reflected from a particle at each 3D position to the camera using predicted surface normals and BRDFs, reducing MLP computation by training only on images illuminated by a single-point light co-located with the camera. Neural Reflectance and Visibility Fields \cite{b58} train an MLP to approximate light source visibility for any input 3D position and 2D incident light direction. The visible light MLP requires only one query per incident light direction, enabling the recovery of a scene's relighting model from images with pronounced shadows and self-occlusion effects. NeLF proposes a human portrait view synthesis and relighting system, using a neural network to predict the light transport field in 3D space and generating human portraits from the predicted neural light transport field in new environmental lighting, achieving simultaneous view synthesis and relighting for given multi-view human portraits.
\subsubsection{reprocessing}\label{5}
In the realm of real-time rendering, post-processing stands as one of the common techniques due to the significant challenges posed by higher resolutions, refresh rates, and more realistic effects. Traditional post-processing methods have relied on linear interpolation techniques such as bilinear interpolation algorithms \cite{b60}, bicubic interpolation algorithms \cite{b61}, among others. However, neural rendering employs neural networks in the post-processing stage to alleviate the burden on the rendering pipeline. Super-resolution and frame interpolation techniques operate at lower resolutions or frame rates within the rendering pipeline and restore the target resolution and frame rate through deep learning methods. Post-processing methods based on deep learning leverage convolutional operations to automatically extract image details, address low-level visual issues, and generate higher quality high-resolution images (e.g., denoising, deblurring). For instance, they use residual networks to enhance inter-layer connections, transmitting low-resolution image feature information to deeper layers to mitigate gradient vanishing and feature loss issues. Additionally, they employ perceptual loss and adversarial loss information from multiple generators in generative adversarial networks to enhance the realism of high-resolution images.

In 2019, NVIDIA introduced the Deep Learning Super Sampling (DLSS) technology, now updated to version 3.0 \cite{b62}, which utilizes deep learning-based super-resolution techniques to construct clearer, higher-resolution images while reducing pixel rendering. The DLSS convolutional autoencoder frame generator receives four input data: the current game frame, the previous game frame, the optical flow field generated by the motion vector accelerator, and game engine data (e.g., motion vectors and depth). For each frame, the DLSS frame generation network determines how to use information from the game motion vectors, optical flow field, and subsequent game frames to generate intermediate frames. Zhang \emph{et al.} \cite{b63} employed a hybrid network architecture of CNN and Transformer to explicitly extract motion and appearance information uniformly, reducing the computational complexity of inter-frame attention while retaining low-level structural details. Additionally, Bi \emph{et al.} \cite{b64} proposed a method to enhance the visual realism of low-quality synthesized images. The authors initially aimed at physics-based rendering, learning to predict accurate shadows in a supervised manner, and then used an improved cycleGAN network \cite{b65} to further enhance the realism of texture and shadows.
\section{Hardware Acceleration}\label{2}
As neural rendering technology becomes widely applied across various domains of real-time rendering, the computational demands for neural rendering increase with the complexity of the neural networks used. Therefore, the exploration of how to better accelerate neural rendering applications at the hardware level has become a research hotspot. Among the 6 papers accepted in the emerging vision/graphics/AR-VR subtopic of the International Symposium on Computer Architecture (ISCA) in 2023, 4 are related to hardware acceleration for neural rendering. This trend underscores the growing importance of hardware acceleration structures for neural rendering.

Hardware architects should systematically consider the following three questions:
\begin{itemize}
    \item The common requirements and performance bottlenecks of neural rendering tasks.
    \item Whether existing general-purpose processors (CPUs), graphics processing units (GPUs), or general-purpose graphics processing units (GPGPUs) can meet the computational and memory access requirements of neural rendering tasks, and whether emerging artificial intelligence processors (neural processing units, NPUs/tensor processing units, TPUs) can provide the operators or special hardware support required for neural rendering tasks.
    \item If the answer is negative, then whether neural rendering requires dedicated hardware acceleration support.
\end{itemize}

\subsection{Common Algorithms}\label{3}
Neural rendering tasks employ various types of neural network algorithms as well as some typical algorithms from traditional computer graphics. This section analyzes the common requirements of the neural rendering techniques introduced in the section on the hardware acceleration.
\subsubsection{Neural Networks}
Since neural rendering is a deep learning-based technology, the deep neural networks utilized by neural rendering algorithms form the core of each algorithm. Table. \MakeUppercase{\romannumeral 3} summarizes the types of neural networks used and the architectural components of the backbone networks employed in the works introduced in Section \ref{2}.
\begin{table*}[!ht]
    \centering
    \caption{The Architecture of Neural Network in Neural Rendering Application}
    \label{table3}
    \scalebox{0.95}{
    \begin{tabular}{|l|l|l|l|}
    \hline
        Neural Network Type & Work & Residual Layer & Concatenation Layer \\ \hline
        MLP & GIRAFFE\cite{b23},neural light transport\cite{b32}, & required & required \\ 
        ~ & neural relighting\cite{b58},NeLF\cite{b59} & ~ & ~ \\ \cline{2-4}  
        ~ & deepSDF\cite{b19},pixelNeRF\cite{b53},IBRnet\cite{b54}, neural radiosity\cite{b66}, & not required & required \\ 
        ~ & common objects in 3D\cite{b55},NeRF\cite{b3},NeRF in the wild\cite{b22}, & ~ & ~ \\ 
        ~ & render net\cite{b28} ,neural voxel renderer\cite{b29},neural volumes\cite{b42} & ~ & ~ \\ 
        ~ & neural sparse voxel fields\cite{b44},kiloNeRF\cite{b45},baking neural radiance fields\cite{b46}, & ~ & ~ \\ 
        ~ & fastNeRF\cite{b47},plenoctrees\cite{b49},autoint\cite{b43},light field networks\cite{b50}, & ~ & ~ \\ 
        ~ & instant neural graphics primitives\cite{b51},neural scene flow fields\cite{b52}, & ~ & ~ \\ 
        ~ & scene representation networks\cite{b56},extracting motion and appearance\cite{b63} ,instant 3D\cite{b67} & ~ & ~ \\ \hline
        CNN & GIRAFFE\cite{b23}, render net \cite{b28},neural voxel renderer\cite{b29}\cite{b32},texture fields\cite{b40}, & required & required \\ 
        ~ & neural reflectance fields\cite{b57},extracting motion and appearance\cite{b63} & ~ & ~ \\ \cline{2-4}  
        ~ & neural volumes\cite{b42},neural point cloud rendering\cite{b31}, & not required & required \\ 
        ~ & neural screen space rendering\cite{b37},deep shading\cite{b38} & ~ & ~ \\ \cline{2-4}  
        ~ & neural control variates\cite{b36},AOGAN\cite{b39},pixelNeRF\cite{b53}, & required & not required \\ 
        ~ & scene representation networks\cite{b56} & ~ & ~ \\ \cline{2-4}  
        ~ & neural light-transport field\cite{b59} & not required & not required \\ \hline
        GAN & deep illumination\cite{b35} & not required & required \\ \cline{2-4}  
        ~ & AOGAN\cite{b39} & not required & not required \\ \hline
        U-Net & deep voxels\cite{b41},neural point-based graphics\cite{b30},deep illumination\cite{b35} & not required & required \\ \hline
        Transformer & IBRnet\cite{b54},common objects in 3D\cite{b55},GeoNeRF\cite{b48}, & required & required \\ 
        ~ & extracting motion and appearance\cite{b63},gen-NeRF\cite{b68} & ~ & ~ \\ \hline
    \end{tabular}
    }
\end{table*}
The residual layer (resblk) is a computational layer where the input of the first unit is linearly stacked with the output of the second unit and then activated, mitigating the vanishing gradient problem.
The concatenation layer (concat) connects multiple input data to one output, achieving data concatenation and allowing for alterations in data dimensions and an increase in the number of channels.

Multilayer perceptrons (MLPs) mainly consist of fully connected layers (FC) and activation layers. Convolutional Neural Networks (CNNs) are primarily composed of convolution layers, activation layers, batch normalization layers (BN), pooling layers, and fully connected layers. Generative Adversarial Networks (GANs) constitute a system comprising two models, a generator (G), and a discriminator (D), with the main network architecture of both being convolutional layers. U-Net, an end-to-end encoder-decoder structure, primarily comprises convolutional layers in the encoder and deconvolutional layers in the decoder. Transformer, fundamentally an encoder-decoder structure, is characterized by the self-attention mechanism as its most crucial component, requiring two matrix-matrix multiplications. Subsequently, the multi-head attention module concatenates and linearly transforms the outputs of H self-attention modules to obtain the final feature matrix. The feature matrix then undergoes normalization residual (add and norm) layers before being fed into a feed-forward neural network (FFN), which consists of a network with 2 FC layers.
\begin{table*}
    \centering
    \caption{The Neural Network Operator Requirement of Neural Rendering Application}
    \label{table4}
    \scalebox{0.94}{
    \begin{tabular}{|l|l|l|l|l|}
    \hline
        Operator & Main Purpose & Operator Features & {\makecell[l]{Hardware Acceleration \\Requirement}} & {\makecell[l]{Neural Rendering \\Process}} \\ \hline
        Convolution/Deconvolution & Feature extraction & High computational demand & Tensor/matrix computation units & Training/Rendering \\ 
        ~ & Data upscaling & ~ & ~ & ~ \\ 
        ~ & Image synthesis & ~ & ~ & ~ \\ \hline
        Matrix-Matrix Multiplication & Attention mechanism & High computational demand & Tensor/matrix computation units & Training/Rendering \\
        ~ & Coordinate transformation & ~ & ~ & ~ \\
        ~ & Feature mapping & ~ & ~ & ~ \\ \hline
        Matrix-Vector Multiplication & Backpropagation & High computational demand & Tensor/matrix computation units & Training/Rendering \\
        ~ & Attention mechanism & ~ & ~ & ~ \\
        ~ & Probability generation & ~ & ~ & ~ \\ \hline
        Pooling/Unpooling & Dimensionality reduction of data & Located on critical path & General computation units & Training/Rendering \\
        (Max, Average) & downsampling & ~ & Execution pipelines & ~ \\ 
        ~ & Increasing receptive field & ~ & ~ & ~ \\ \hline
        Element-wise Operations  & Residual modules, Bias & High memory access demand & General computation units & Training/Rendering \\ 
        (Multiplication, Addition) & ~ & ~ & Data transfer units & ~ \\ \hline
        Activation Functions  & Probability generation & Located on critical path & Table lookup units & Training/Rendering \\ 
        {\makecell[l]{(softmax, ReLU, tanh, \\sigmoid, etc.)}} & Function approximation & ~ & Execution pipelines & ~ \\ \hline
        Normalization & Gradient vanishing & Located on critical path & General computation units & Training/Rendering \\ 
        ~ & ~ & ~ & Execution pipelines & ~ \\ \hline
    \end{tabular}
    }
\end{table*}
The primary common operators in deep neural networks, as shown in Table. \MakeUppercase{\romannumeral 4}, include convolution/deconvolution, matrix-matrix multiplication, matrix-vector multiplication, pooling/unpooling, element-wise multiplication/addition operations, activation functions, and normalization. Operators such as convolution and matrix-matrix multiplication represent the most computationally intensive operations in neural networks, necessitating specialized hardware units for acceleration, such as systolic arrays, dot products, and multi-level accumulation trees. These hardware units demonstrate superior acceleration effects compared to general-purpose processors utilizing single instruction multiple data (SIMD) and graphics processors utilizing single instruction multiple thread (SIMT) data parallelism. For operations like pooling/unpooling and activation functions, although they do not contribute significantly to computational load, their inherently serial execution characteristics occupy the program's critical path runtime, thus requiring dedicated hardware units for acceleration, such as parallel look-up tables, transcendental function/interpolation special operation units. As for element-wise operations, although they do not entail significant computational load, they demand high memory bandwidth. While not mandating the design of specialized hardware computing acceleration units, they necessitate optimized memory access paths to enhance operation efficiency, such as designing dedicated on-chip network structures or optimizing memory units to support multiple data transfer modes.

\subsubsection{Ray Marching and Volume Rendering}
Ray marching refers to the incremental progression of a ray from its origin, halting at each step to perform computations before continuing until it approaches the endpoint. This method is commonly employed in implicit surface, point cloud, or voxel rendering. Many neural rendering tasks, particularly those akin to Neural Radiance Fields (NeRF), utilize ray marching for rendering, replacing traditional ray tracing or rasterization. Table. \MakeUppercase{\romannumeral 5} enumerates the relevant works employing ray marching techniques.

\begin{table*}[!ht]  
\caption{\textbf{The Ray Marching in Neural Rendering Application}}  
\label{table5}  
\scalebox{1.1}{
\begin{tabular}{|l|l|l|l|l|}  
\hline  
\textbf{Work} & \textbf{Type} & \textbf{Purpose} & \textbf{Neural Rendering Process} \\  
\hline  
Neural Volumes \cite{b42} & Accumulation-Ray Marching & Render Voxel & Rendering \\ \hline   
NeRF\cite{b3} and its variants\cite{b46}--\cite{b55} & Ray-marching-Ray Stepping & Render Voxel & Rendering \\  \hline 
Common Objects in 3D \cite{b55} & Radiance Absorption-Ray Stepping & Render Transparency Field & Rendering \\ \hline  
Scene Representation Networks \cite{b56} & Differential-Ray Stepping & Render Voxel & Training \\ 
\hline

\end{tabular}  
}
\end{table*}  
In general, the ray marching in neural rendering tasks is mainly executed in three steps: first, determining the number of rays to be generated; then, establishing the stepping points for the current ray propagation; and finally, querying an MLP network based on the coordinates to calculate the color and writing it back to the pixel. The core computation in ray marching is the GetDist function, which takes the coordinates of a point in space as input and returns the distance from this point to the nearest object in space. Typically, the SDF function is used to calculate the distance from a point in space to a spherical surface, as shown in \eqref{eq3}:
\begin{equation}
    \Phi_{\text {circle }}(\boldsymbol{x})=\|\boldsymbol{x}-\boldsymbol{c}\|-r
    \label{eq3}
\end{equation}
In \eqref{eq3}, $x$ represents the coordinates of a point in space, $c$ denotes the center coordinates of the sphere, and $r$ represents the radius of the sphere. When the function value is greater than 0, it indicates that the coordinates are outside the various objects in the scene. When the function value is less than 0, it indicates that the coordinates are inside the various objects in the scene. When the function value equals 0, it indicates that the coordinates are on the surfaces of the various objects in the scene. Because the calculation of the 3D coordinates in \eqref{eq3} can be geometrically transformed into a vector modulus problem, the calculation of ray marching can be transformed into vector multiplication, square root, and scalar addition. Hence, the computational demand for ray marching is a general computational unit, without the need for specialized hardware acceleration units.

Neural rendering tasks, especially NeRF-like tasks, require the use of traditional volume rendering calculations. This involves computing the final color of the sample point by applying the colors and density results obtained from querying the MLP network through the volume rendering formula. The discrete volume rendering calculation formula is represented as  \eqref{eq4}:

\begin{equation}
    \hat{C}(r)=\sum_{k=1}^{N} T_{k}\left(1-\exp \left(-\sigma_{k}\left(t_{k+1}-t_{k}\right)\right)\right) c_{k}
\label{eq4}
\end{equation}
\begin{equation}
T_{k}=\exp \left(-\sum_{j=1}^{k-1} \sigma_{j}\left(t_{j+1}-t_{j}\right)\right)
    \label{eq5}
\end{equation}
From \eqref{eq4}, it can be inferred that the computational requirements for volume rendering consist of general computational units and lookup table units, aligning with the hardware acceleration unit requirements for neural network computations.
\subsubsection{Other Considerations}
In addition to neural networks, ray marching, and volume rendering calculations, linear interpolation is also a common requirement. Table. \MakeUppercase{\romannumeral 6} presents the usage of interpolation calculations in neural rendering work. 
\begin{table*}
    \centering
    \caption{\textbf{The Interpolation in Neural Rendering Application}}
    \label{table6}
    \begin{tabular}{|l|l|l|l|l|}
    \hline
        Work & Interpolation Type & Data Type & Purpose & Neural Rendering Process \\ \hline
        deep voxels\cite{b41} & 3-linear & Voxel Feature & Voxel Feature Interpolation & Rendering \\ \hline
        Instant NGP\cite{b51} & d-linear & Voxel & Hash Encoding Interpolation & Training \\ \hline
        deep voxels\cite{b41},neural point-based graphics \cite{b30},& Bi-linear & Feature & Deconvolution Interpolation & Rendering \\ 
        deep illumination \cite{b35} & ~ \\ \hline
        Traditional Post-processing & Bi-linear/Bi-cubic & Pixel & Pixel Interpolation & Rendering \\ \hline
    \end{tabular}
\end{table*}
Linear interpolation calculations can be applied not only in end-to-end rendering pipelines for neural rendering but also in neural rendering pipelines combined with traditional post-processing pipelines. The computational demand for linear interpolation involves general computational units and special function units, aligning with the hardware acceleration requirements of pooling layers and pointwise operation layers in ray marching and neural networks. Furthermore, the introduction of multi-resolution hash encoding in Instant NGP \cite{b51} has proven to accelerate the training process of NeRF-like neural rendering tasks. Hash encoding issues can be transformed into hardware look-up tables, aligning with the hardware acceleration requirements of activation function operators in neural networks.
\subsection{Requirements Analysis}
Based on the common requirements discussed in Section \ref{3}, this section separately discusses the neural rendering task in terms of training and inference processes.
\subsubsection{Training Process}\label{4}
The training process of the neural rendering forward task is fundamentally similar to the training processes of other deep learning applications. It involves the use of the backpropagation algorithm for training, with data types being single-precision floating-point or half-precision floating-point data. The general steps include initially pretraining the model in a supervised or unsupervised manner on a large dataset and then fine-tuning the pretrained model to adapt to specific downstream tasks on a test set. The training process of the neural inverse rendering task slightly differs from training processes based on deep learning applications.
\Figure[t!](topskip=0pt, botskip=0pt, midskip=0pt)[width=1\linewidth]{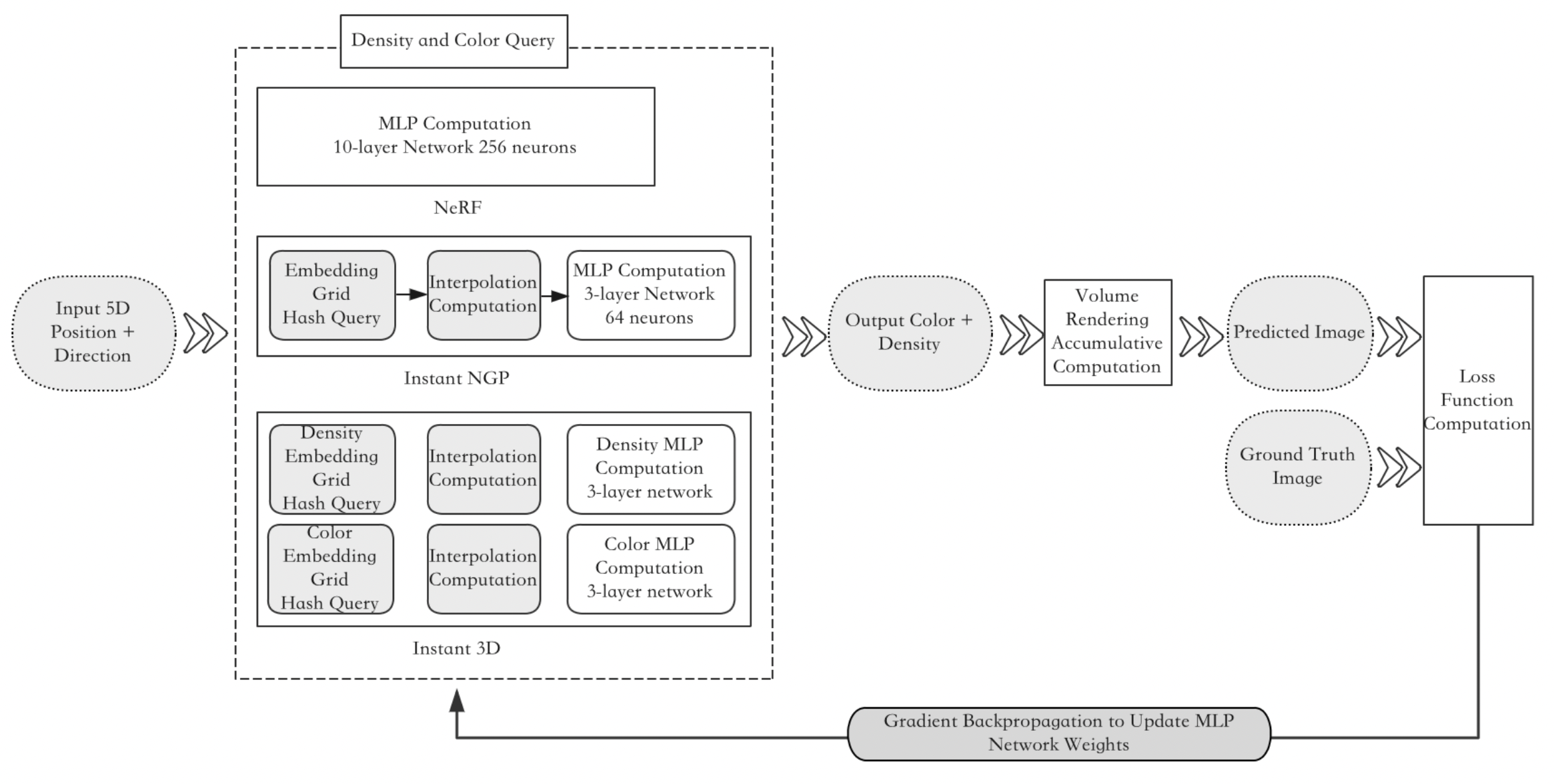}
{\textbf{The render process versus of NeRF, instant NGP and instant 3D.}\label{fig7}}
Fig. \ref{fig7} presents a comparison of the training processes for NeRF \cite{b3}, Instant NGP \cite{b51}, and Instant 3D \cite{b67}. In the original NeRF framework, training for a single scene typically requires convergence over 100,000 to 300,000 iterations. Each batch processing scale, i.e., the number of ray sampling points per pixel, amounts to 192 points/pixel multiplied by 4,096 pixels. Therefore, each iteration necessitates 786,432 MLP queries, resulting in a training computational load approaching $10^{17}$ . Utilizing a single NVIDIA V100 GPU, the training duration typically requires approximately 1 to 2 days. Instant NGP achieves comparable training effects to the original NeRF in a matter of 5 seconds on the same hardware platform through algorithmic optimizations. Li \emph{et al.} \cite{b67} expedited the NeRF training process through a collaborative design of software and hardware. They proposed the Instant 3D algorithm based on Instant NGP and its corresponding hardware acceleration structure. By decoupling the color and density branches of the embedded grid without compromising reconstruction quality, they compressed NeRF computations to enhance training efficiency. Consequently, they achieved a training time of 1.6 seconds and a training power consumption of 19,000 watts per scene, with virtually unchanged quality on AR/VR devices. This represents a reduction in training time by 41 to 248 times compared to traditional NeRF-like applications. Subsequent work on neural rendering training has predominantly adopted hash encoding as a means of training acceleration.

Numerous comprehensive works have already addressed training methodologies for various types of neural networks. Zhou \emph{et al.} \cite{b71} discussed supervised and unsupervised learning methods for Convolutional Neural Networks (CNNs). Wang \emph{et al.} \cite{b72} examined the primary challenges faced in single-machine and distributed training, providing an overview of optimization algorithms and representative achievements across research branches. Additionally, Zhang \emph{et al.} \cite{b73} and Yin \emph{et al.} \cite{b74} respectively explored the training aspects of Transformer network architectures in image and video tasks. Therefore, this paper does not delve into the general training issues of neural networks.

\subsubsection{Rendering Process}\label{1}
The rendering process for neural rendering tasks closely resembles the inference process of other deep learning applications. The data types involved are typically half-precision floating point or integer. The general steps include input encoding, neural network queries, and finally, pixel output generation. The rendering process primarily employs neural networks for feature extraction, coordinate transformation, and image synthesis functions. Algorithms utilizing MLP coordinate projection combined with CNN generation networks as the rendering framework require substantial convolutional and matrix computational power. For instance, Render Net employs a 1-layer 32×32×512 scale MLP network for coordinate projection and an 8-layer CNN network for generating 512×512×3 RGB images. In neural rendering algorithms using U-Net as the generation model, the inference process necessitates the use of a pre-trained U-Net for image generation, requiring significant convolutional and general computational power. For instance, as depicted in Fig. \ref{fig8}, the rendering generation process of Neural Light Transport utilizes downsampling with a stride of 2 using convolutions, followed by linear interpolation for upsampling, and incorporates 3 skip connections (original U-Net has 4 skip connections) to merge information for image synthesis. Skip connection layers can also be considered as a form of residual layer, fundamentally involving element-wise operations. Additionally, the skip connection and concatenation layers in Fig. \ref{fig8} require storing input images and two intermediate feature maps, thereby increasing memory access overhead.
\Figure[t!](topskip=0pt, botskip=0pt, midskip=0pt)[width=1\linewidth]{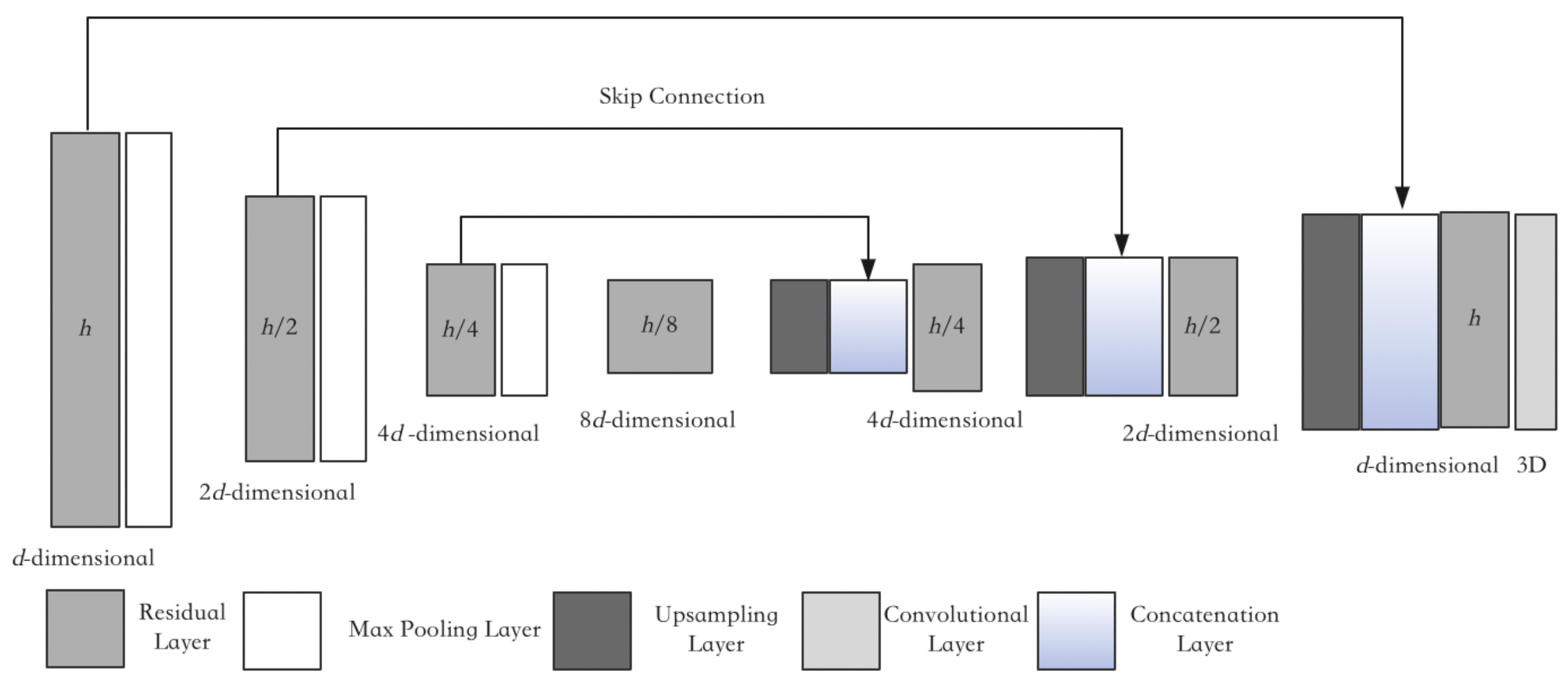}
{\textbf{The network architecture of image synthesis layer.}\label{fig8}}
In summary, the hardware acceleration requirements for the forward rendering process in neural rendering applications encompass matrix operations, convolutional operations, and general computational capabilities. The specific computational and memory requirements for different tasks are determined by the voxel count in the target scene and the resolution of the rendered images.
\Figure[t!](topskip=0pt, botskip=0pt, midskip=0pt)[width=1\linewidth]{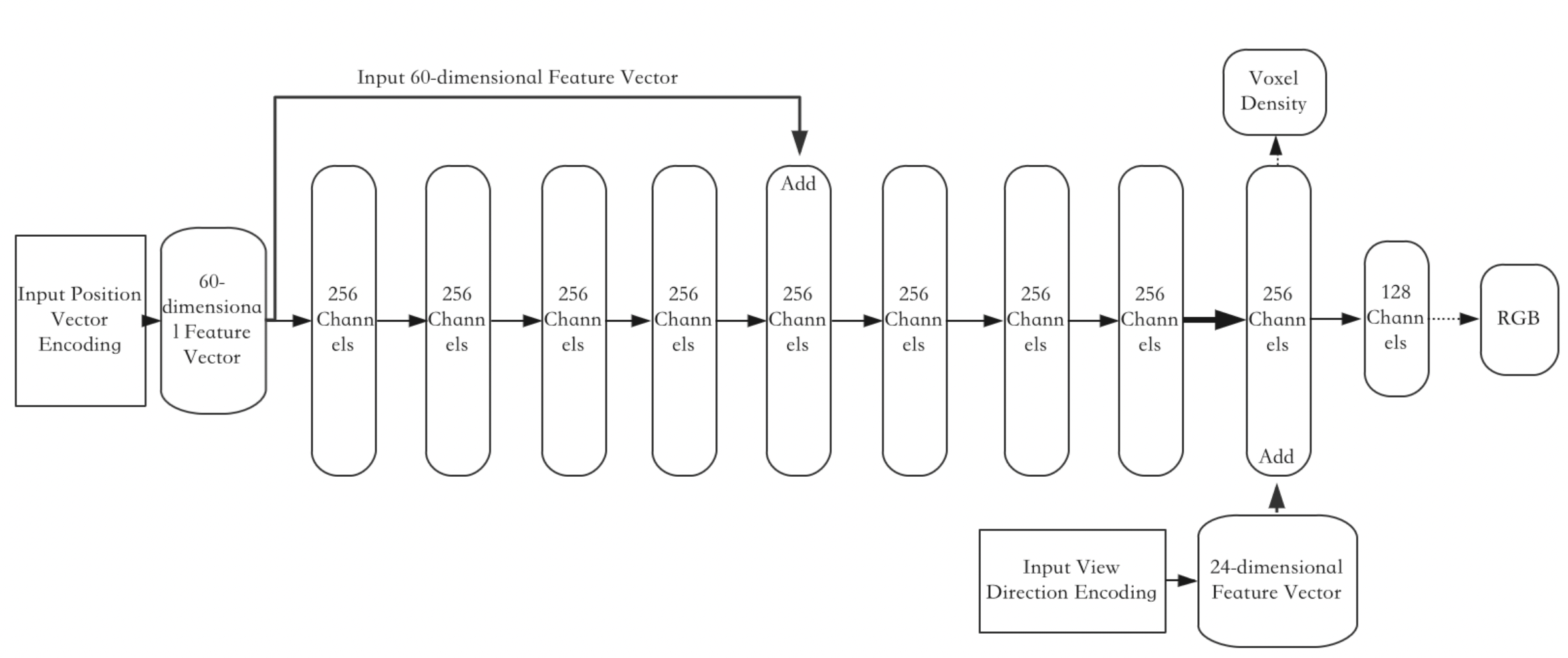}
{\textbf{The FC network architecture of NeRF.}\label{fig9}}

In the neural inverse rendering application, the rendering process primarily employs neural networks for input encoding, scene querying, and ray marching algorithms for ray tracing. Taking the original NeRF as an example, the authors selected 64 points per ray (for coarse scenes) and 128 points per ray (for fine scenes), totaling 192 ray marching points. For each ray marching point, the MLP network, as depicted in Fig. \ref{fig9}, first encodes the 3D coordinate vector into a 60-dimensional vector for input and processes it through an MLP network comprising 8 FC layers (where the 5th layer requires re-inputting the 60-dimensional vector to enhance coordinate information), with each layer having 256 channels. This process yields an output of a 256-dimensional feature vector and $\sigma$. Subsequently, the 256-dimensional feature vector is concatenated with a 24-dimensional vector generated from the input encoding of the camera viewing direction and is then fed into an FC layer with 256 channels followed by an FC layer with 128 channels to compute the RGB color. To achieve high-quality rendering, the authors set the number of rays per image to be 762k, requiring 150 to 200 million ray marching and MLP network computations to render a single frame, taking 30s on an NVIDIA V100 GPU to render one frame.

FastNeRF \cite{b47} splits the sequential coordinate encoding MLP network and the ray direction MLP network in NeRF into two parallel MLP networks for accelerated rendering. One network generates d-dimensional neural radiance volume outputs using an 8-layer position-aware MLP network, while the other produces 1-dimensional directional feature outputs using a 4-layer direction-aware MLP network. Experimental results on an NVIDIA RTX 3090 GPU show that FastNeRF is 3000 times faster than the original NeRF in rendering speed. Notably, the cache occupancy of FastNeRF's MLP network weights is 54GB, which can be placed in the GPU's on-chip cache to enhance rendering speed. MobileNeRF \cite{b75} decomposes NeRF's MLP network into an encoder (first 8 layers) and a decoder (last 2 layers), pre-storing the output of the encoder and only executing the inference calculation of the decoder to achieve real-time NeRF neural rendering on mobile devices. However, MobileNeRF also faces several issues, such as high requirements for the training dataset (otherwise, numerous voids may appear) and lengthy training times (requiring training in 3 steps).

The R2L model \cite{b76} replaces NeRF with a neural light field NeLF to accelerate rendering speed. Due to the model’s direct output of RGB and the absence of volumetric density learning and opacity synthesis steps, predicting pixel colors only requires a single forward pass of the rays, making R2L significantly faster in rendering compared to NeRF. However, NeLF is more challenging to train compared to NeRF. To address this, the authors employed an 88-layer deep MLP architecture with residual layers as the mapping function, resulting in longer rendering delays. Additionally, NeLT requires NeRF to generate pseudo datasets to aid in training. MobileR2L \cite{b77}, based on the R2L model, proposed an architecture optimization tailored for mobile devices. In contrast to MobileNeRF, it does not require pre-storing intermediate data, making it more suitable for hardware platforms with limited computational capabilities. However, the method proposed in the paper leads to significant memory issues when handling large ray sizes.
\Figure[t!](topskip=0pt, botskip=0pt, midskip=0pt)[width=1\linewidth]{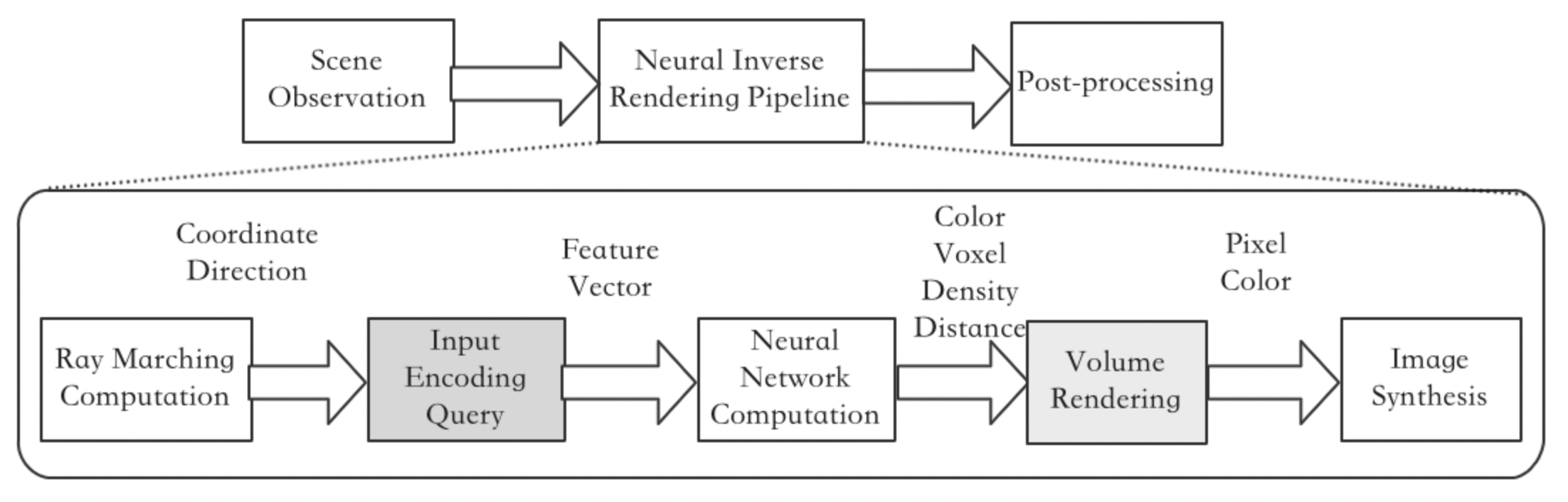}
{\textbf{Pipeline of inverse rendering.}\label{fig10}}
In summary, Fig. \ref{fig10} illustrates the pipeline of the neural inverse rendering task, where each sampling point on every ray independently executes this pipeline, thus naturally possessing high parallelism that can be accelerated by hardware. Hardware acceleration requirements for the inverse rendering process in neural rendering applications include input encoding queries, ray marching, matrix operations, convolution operations, and general computational capabilities. Compared to forward rendering, inverse rendering requires higher computational power and storage. The ray marching, encoding queries, volume rendering, neural network computations, rendering resolution, and the number of generated rays are closely related. Obtaining high-quality images entails significant computational and memory overhead. Consequently, achieving faster input encoding queries and MLP network computations becomes a key focus of hardware acceleration.

\subsubsection{Bottleneck Analysis}
From Section \ref{4}, it can be inferred that the training processes of neural forward rendering and inverse rendering applications exhibit characteristics of high computational load and demanding memory bandwidth. In comparison to inverse rendering applications, neural forward rendering tasks demonstrate lower computational and memory requirements. For instance, considering the deep illumination model \cite{b35} which generates 3-channel 256×256 pixel images using a 16-layer neural network, a single NVIDIA P500 GPU requires 3 hours. Conversely, for neural inverse rendering tasks such as those exemplified by NeRF \cite{b3}, a single NVIDIA V100 GPU may require 1 to 2 days for a single scene. Therefore, the training processes for neural rendering tasks, especially for inverse rendering applications, necessitate substantial computational and storage resources. Typically, the training of neural rendering processes is performed offline, often running on hardware platforms at the level of cloud data centers or server-grade hardware platforms, as the computational and storage capabilities of device-level hardware platforms are insufficient to support such training tasks.
\Figure[t!](topskip=0pt, botskip=0pt, midskip=0pt)[width=1\linewidth]{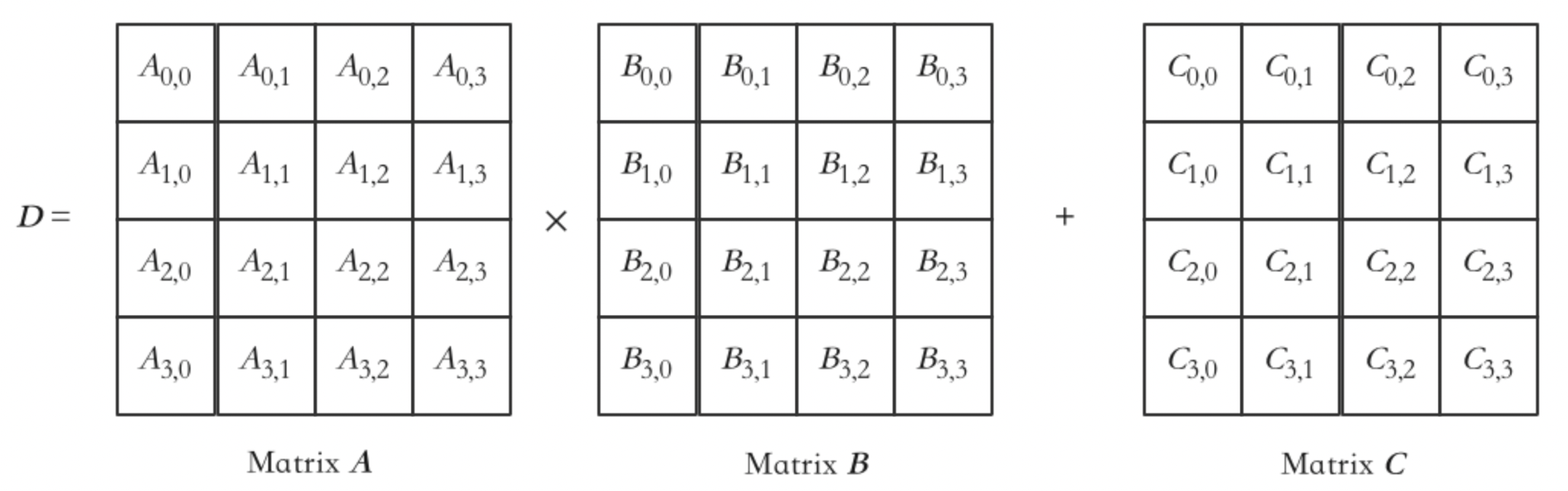}
{\textbf{The MAC operation in tensor core.}\label{fig11}}
From the section on rendering process of requirements analysis in hardware acceleration, it can be observed that in the rendering process of neural forward rendering and inverse rendering applications, neural forward rendering applications typically achieve real-time rendering at high resolutions, while inverse rendering applications highlight performance bottlenecks. Muhammad \emph{et al.} \cite{b78} evaluated four algorithms (NeRF, NSDF, GIA, NVR) on an NVIDIA RTX3090 GPU (35.58 TFLOPS@FP16) using a multi-resolution hash encoding method \cite{b51}. Rendering a single frame of size 1920 × 1080 required total runtimes of 231 ms, 27.87 ms, 2.12 ms, and 6.32 ms for the respective algorithms, which is deemed unacceptable for real-time rendering. Notably, the subtasks of embedded grid hash encoding and MLP queries are the most time-consuming performance bottlenecks in neural rendering applications, jointly accounting for approximately 70\% of the total duration. Fu \emph{et al.} \cite{b68} assessed NeRF synthesis and deep voxels tasks, reporting that they required 28 s and 13 s, respectively, on an NVIDIA RTX 2080Ti desktop-grade GPU, whereas on an NVIDIA Jetson TX2 edge-grade GPU, they required 1000 s and 380 s, respectively. Here, the runtimes of the subtasks involving MLP queries and ray transformer calculations constituted approximately 70\% to 90\% of the total runtime. Therefore, embedded grid hash encoding, MLP queries, and ray transformer calculations are the primary targets for hardware platform acceleration.
\Figure[t!](topskip=0pt, botskip=0pt, midskip=0pt)[width=1\linewidth]{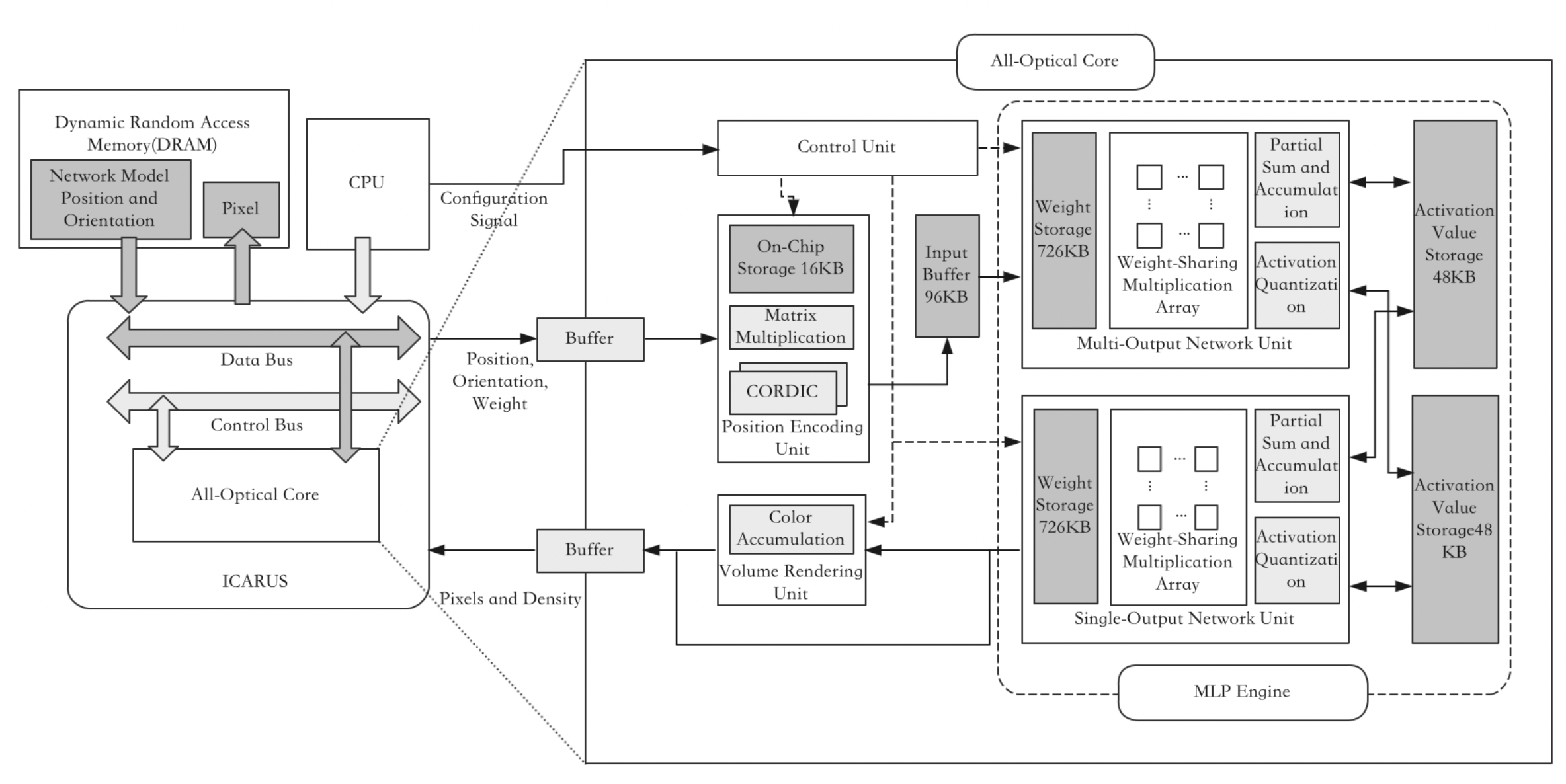}
{\textbf{The architecture of ICARUS.}\label{fig12}}
\Figure[t!](topskip=0pt, botskip=0pt, midskip=0pt)[width=1.0\linewidth]{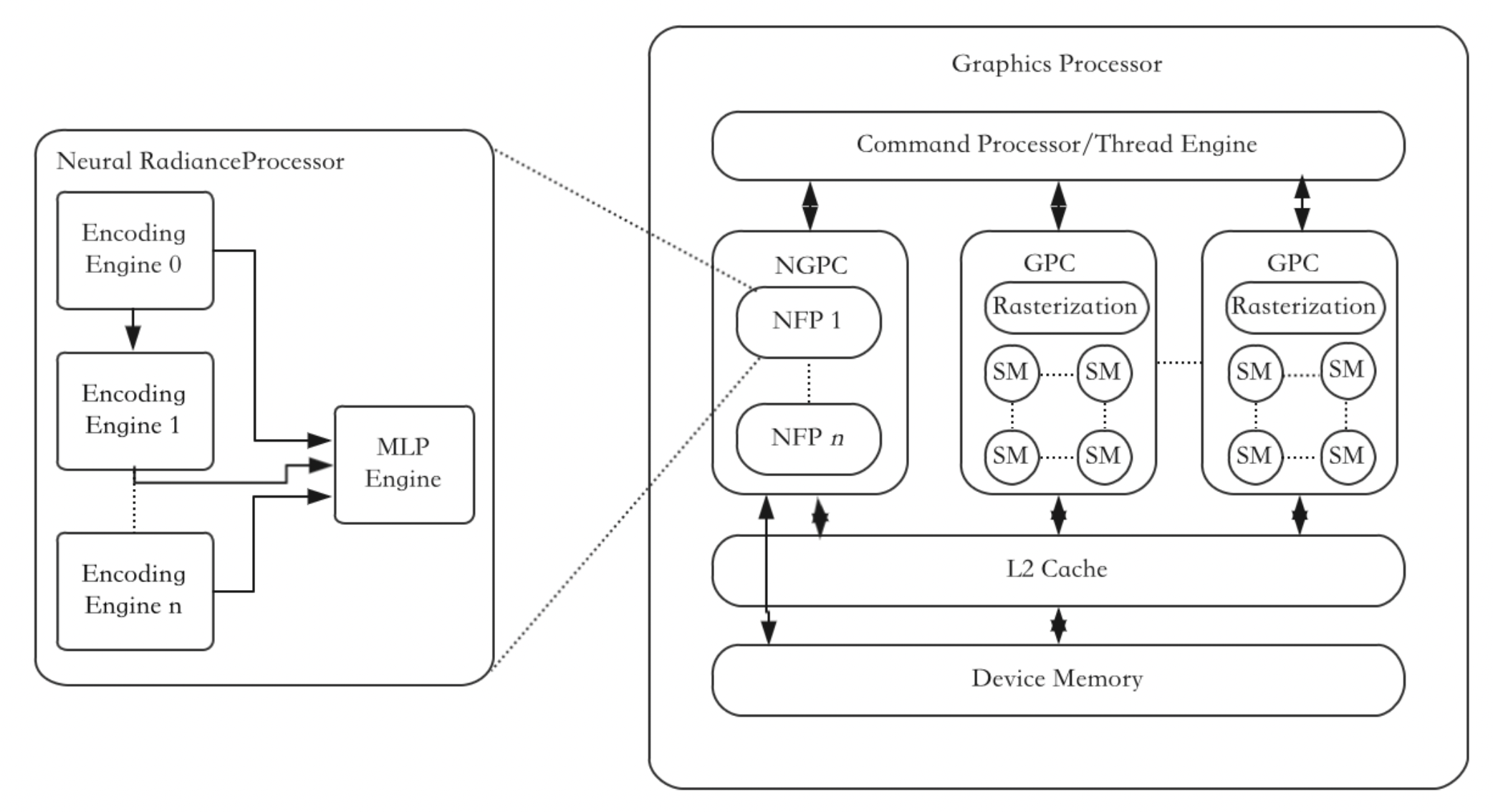}
{\textbf{The architecture of NGPC.}\label{fig13}}
\Figure[t!](topskip=0pt, botskip=0pt, midskip=0pt)[width=1\linewidth]{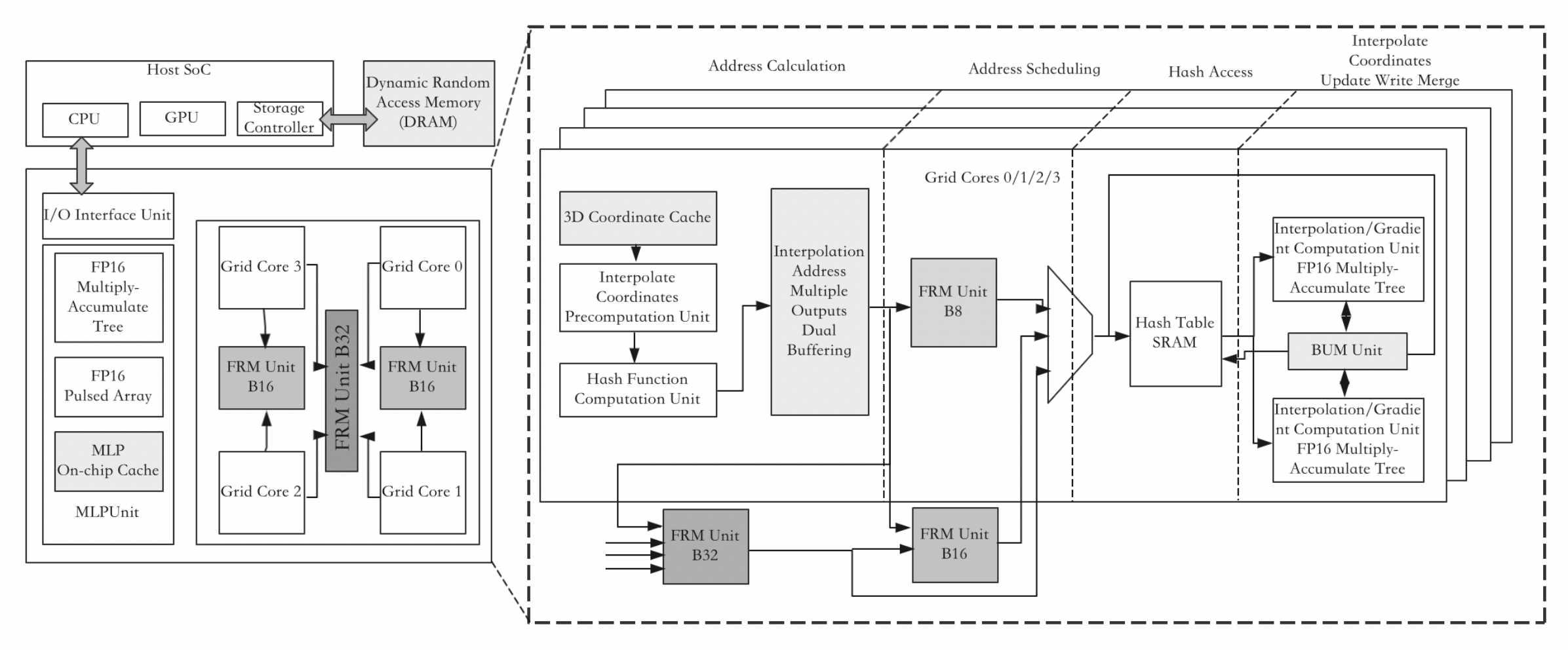}
{\textbf{The architecture of Instant 3D accelerator.}\label{fig14}}
Li \emph{et al.} \cite{b67} designed the Instant 3D accelerator for 3D reconstruction and neural rendering tasks akin to NeRF. The Instant 3D accelerator, illustrated in Fig. \ref{fig14}, primarily consists of three parts: MLP units, grid cores, and I/O interfaces. The authors mapped the embedded grid hash queries and MLP queries from the Instant 3D algorithm to the Instant 3D accelerator for hardware pipelining acceleration. Additionally, they designed a forward read mapper (FRM) to merge multiple memory read requests and a backward update merger (BUM) to combine multiple grid updates into a single update, thus enhancing the on-chip SRAM array utilization, minimizing SRAM write counts and power consumption, and supporting various grid sizes required by the Instant 3D algorithm. Fu \emph{et al.} \cite{b68} devised the Gen-NeRF accelerator for real-time, generalizable NeRF-like tasks. The overall architecture of the Gen-NeRF accelerator, as depicted in Fig. \ref{fig15}, mainly comprises a software-defined rendering engine, workload scheduler, and on-chip memory. The authors employed epipolar geometry inference to expedite the target workload and designed the rendering engine architecture to optimize the software-to-hardware data flow mapping, maximizing data reuse among rays.
\Figure[t!](topskip=0pt, botskip=0pt, midskip=0pt)[width=1\linewidth]{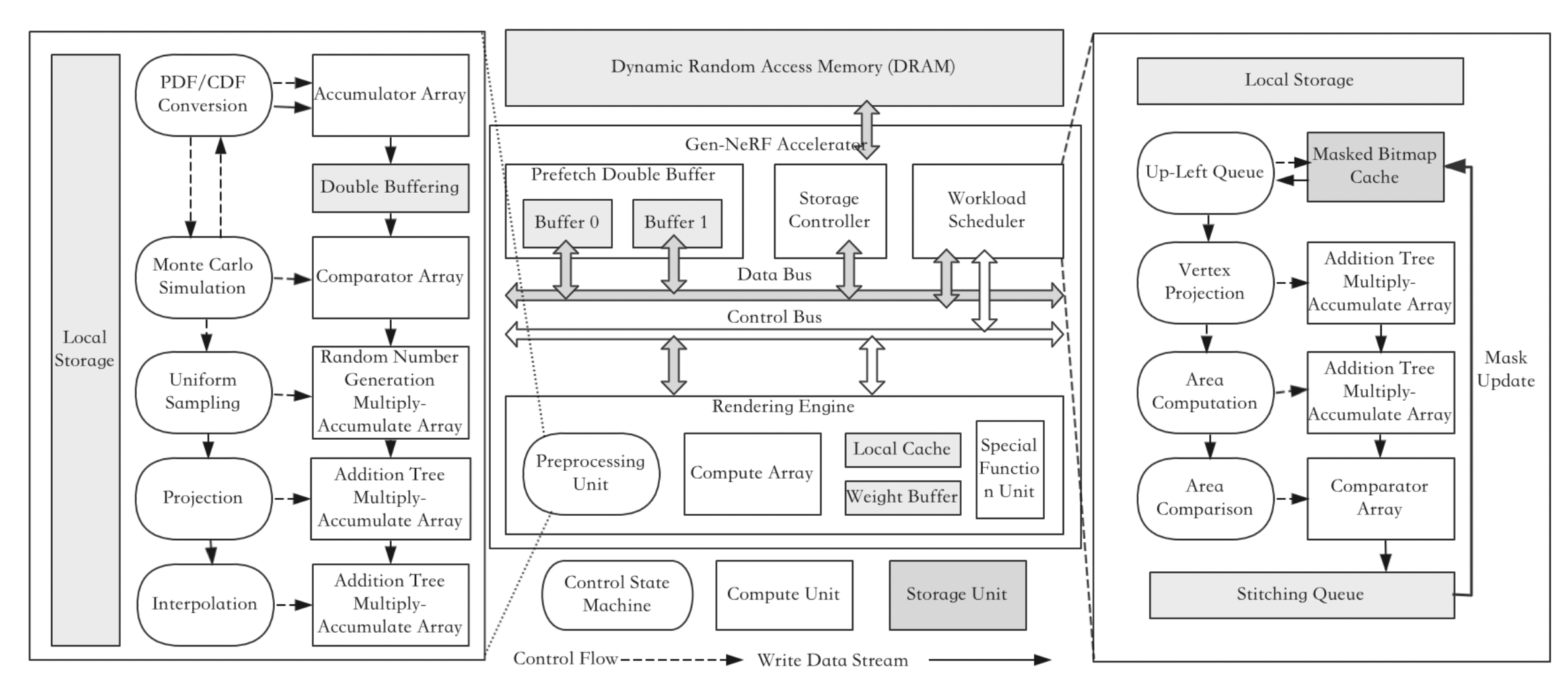}
{\textbf{The architecture of Gen-NeRF accelerator.}\label{fig15}}
\subsection{Platform Status}
Common hardware processing platforms for graphic applications include general-purpose processing units, graphics processing units, and domain-specific architectures (DSAs) represented by tensor processors and neural processors \cite{b79}. Currently, these hardware processing platforms all provide support for neural networks and are applicable to neural rendering applications. This section respectively introduces their current status in neural network acceleration and conducts an adaptability analysis for neural rendering applications.
\subsubsection{General-Purpose Processor (CPU)}
As the most widely used hardware processing platform, general-purpose processors (CPUs) mainly focus on three aspects of hardware support for neural network acceleration: 
\begin{table*}[!ht]
    \centering
    \caption{\textbf{The Acceleration for NN in CPU}}
    \label{table7}
    \scalebox{1.07}{
    \begin{tabular}{|l|l|l|l|l|}
    \hline
        Platform & Instruction Set & Data Type & Hardware Unit & Scene Orientation \\ \hline
        IBM Power10\cite{b80} & Power-ISA 3.1 MMA Instruction Extension & FP16 & MMA Computing Unit & Inference/Training \\ \hline
        Tesla DOJO\cite{b81} & 64B Specialized Instruction Set & BF16 & Vector/Matrix Coprocessing Unit & Inference/Training \\ 
        ~ & ~ & CFP8 & Scatter/Gather Unit & ~ \\ \hline
        AMD EPYC & Scatter/Gather Instruction & BF16 & Mul MAC Unit & Inference/Training \\ 
        ~ & BF16 Instruction & FP16 & ~ & ~ \\
        ~ & VNN (Vector Neural Network) Instruction & INT8 & ~ & ~ \\ \hline
        NVIDIA Grace & BF16 Instruction & BF16 & MatMul Unit & Inference/Training \\ 
        ~ & Scatter/Gather Instruction & INT8 & ~ & ~ \\ \hline
        Intel Xeon 4th Gen & TMUL Instruction & BF16 & AMX Unit & Inference/Training \\ 
        ~ & VNN Instruction & INT8 & ~ & ~ \\ \hline
    \end{tabular}
    }
\end{table*}
\begin{itemize}
    \item increasing neural network extension instruction sets;
    \item introducing neural network-specific data types;
    \item incorporating neural network-specific acceleration components. Table. \MakeUppercase{\romannumeral 7} presents the acceleration support for neural networks across different CPU platforms.
\end{itemize}

The IBM Power10 processor \cite{b80} introduces a new set of Matrix Multiply Assist (MMA) instructions and corresponding MMA compute units in the latest Power-ISA 3.1 version. It directly performs numerical linear algebra operations on small matrices while supporting half-precision data operation instructions. These instructions can accelerate compute-intensive kernels such as matrix multiplication, convolution, discrete Fourier transform, and artificial intelligence workloads including neural networks. When facing neural network workloads, the addition of MMA units in the POWER10 processor can improve energy efficiency at the single-core level by 2.6 times compared to POWER9.

Dojo \cite{b81} introduces the microarchitecture of Tesla's high-throughput general-purpose processor for AI training. The instruction width of the general processing cores in the DOJO processing node is 64 b, while the instruction width of the vector/matrix coprocessor is 64 B. Coprocessor instructions are sent to the vector scheduling module for processing through the scalar scheduling module interface. The memory access path width for the general processing cores is 8B×2, while for the vector/matrix coprocessor, it is 64B×3. The general processing cores and coprocessors use SRAM space instead of data cache for data interaction. Therefore, in DOJO, the general processing cores are only responsible for general address calculation and logical computation, while the calculation tasks for neural network workloads are handled by the coprocessor.

Both AMD and Intel have respectively added the Vector Neural Network (VNN) instruction set and corresponding hardware acceleration units for accelerating neural network workload computation in their latest EPYC Milan processors and Xeon 4th Gen processors. They have also added Scatter/Gather memory access instruction extensions to accelerate memory access for neural network workloads. Additionally, AMD, Intel, and ARM have all added support for the BF16 data type in their latest CPUs.

In both server-level and desktop/mobile mainstream CPU platforms, architectural-level optimizations for neural network applications have been implemented. The addition of neural network compute units and corresponding instruction sets to CPU platforms has three main benefits: 
\begin{itemize}
    \item no impact on other parts of the original architecture, minimal impact on the CPU microarchitecture;
    \item CPUs can reduce the power consumption impact of neural network units through gated clocks or dynamic switching; 
    \item they can inherit the original programming model and possess a certain level of versatility.
\end{itemize}
\subsubsection{Graphics Processing Unit (GPU)}
The Graphics Processing Unit (GPU) or General-Purpose Graphics Processing Unit (GPGPU) possesses significantly greater computational and storage resources than the CPU, along with powerful floating-point calculation capabilities. Their architecture is exceptionally well-suited for handling large-scale, highly parallel, computation-intensive tasks. Since its inception in 1998, the GPU has undergone rapid development, evolving from a dedicated platform for graphics tasks to encompassing hardware processing platforms for high-performance general computing, graphics processing, and deep learning computations. Currently, the GPU/GPGPU stands as the most widely deployed hardware platform for neural network applications, with many deep learning frameworks based on GPU platform interfaces. Hardware support for accelerating neural networks primarily focuses on the following four aspects:
\begin{itemize}
    \item Chip architectures suitable for neural network computation.
    \item Introduction of dedicated data types for neural networks.
    \item Inclusion of specialized acceleration components for neural networks.
    \item Addition of dedicated memory access units for neural networks.
\end{itemize}
\begin{table*}[!ht]
    \centering
    \caption{\textbf{The Acceleration for NN and Ray in GPU/GPGPU}}
    \label{table8}
    \scalebox{1.02}{
    \begin{tabular}{|l|l|l|l|l|l|l|}
    \hline
        Platform & {\makecell[l]{Peak Compute\\/FLOPS}} & Data Type & {\makecell[l]{Hardware \\Accelerator\\ Unit}} & {\makecell[l]{Memory Access \\Accelerator Unit}} & {\makecell[l]{Ray Tracing \\Accelerator Unit}} & {\makecell[l]{Scene \\Orientation}} \\ \hline
        NVIDIA H100\cite{b82} & 500T@TF32 & TF32 & Tensor Core & {\makecell[l]{Tensor Memory \\Accelerator Unit}} & None & {\makecell[l]{Training/Inference/\\High Performance}} \\ 
        ~ & 1000T@FP16 & FP16 & ~ & TMA & ~ & ~ \\ 
        ~ & 1000T@BF16 & BF16 & ~ & ~ & ~ & ~ \\ 
        ~ & 2000T@FP8 & FP8 & ~ & ~ & ~ & ~ \\ \hline
        AMD MI250\cite{b83} & 383T@FP16 & FP16 & Matrix Core & None & None & {\makecell[l]{Training/Inference/\\High Performance}} \\
        ~ & 383T@BF16 & BF16 & ~ & ~ & ~ & ~ \\ 
        ~ & 383@INT8 & ~ & ~ & ~ & ~ & ~ \\ \hline
        {\makecell[l]{Intel Ponte Vecchio \\XMX\cite{b84}}}& 419T@TF32 & TF32 & Matrix Engine & {\makecell[l]{Control Flow \\Processing \\Prefetch Unit}} & None & {\makecell[l]{Training/Inference/\\High Performance}} \\ 
        ~ & 839T@FP16 & FP16 & ~ & CSP & ~ & ~ \\ 
        ~ & 839T@BF16 & BF16 & ~ & ~ & ~ & ~ \\ 
        ~ & 1678@INT8 & ~ & ~ & ~ & ~ & ~ \\ \hline
        Biren BR100\cite{b85} & 512T@TF32 & TF32 & Vector Unit & {\makecell[l]{Tensor Data Storage \\Accelerator Unit}} & None & {\makecell[l]{Training/\\Inference}} \\ 
        ~ & 1024T@BF16 & BF16 & ~ & TDA & ~ & ~ \\ 
        ~ & 2048T@INT8 & ~ & ~ & ~ & ~ & ~ \\ \hline
        {\makecell[l]{NVIDIA GeForce \\RTX 4090\cite{b86}}} & 660.6T@FP8 & FP8 & Tensor Core & None & Ray Tracing Unit & Inference/Graphic \\ 
        ~ & 191T@ray & ~ & ~ & ~ & ~ & ~ \\ 
        {\makecell[l]{AMD RADEON™ \\RX 6950 XT\cite{b87}}} & 47.31T@FP16 & FP16 & Matrix Core & None & Ray Tracing Accelerator & Inference/Graphic \\ \hline
    \end{tabular}
    }
\end{table*}
Table. \MakeUppercase{\romannumeral 8} provides an overview of the acceleration support for neural networks across different GPU/GPGPU platforms.

NVIDIA, starting from the Kepler architecture, made substantial alterations to its GPU architecture, transitioning from a graphics processing-oriented architecture to a general-purpose computing architecture. In the Volta architecture, NVIDIA introduced tensor cores for the first time to accelerate neural network tasks. Each Tensor Core can perform mixed-precision operations for 4×4×4 matrix multiplication and accumulation, as depicted in Fig. \ref{fig11}. In its latest H100 GPU, the peak performance for half-precision floating point (FP16) has reached 1000 TFLOPS. Furthermore, NVIDIA has sequentially incorporated dedicated data types for neural network training/inference, such as TF32, BF16, FP8, and introduced Tensor Memory Access (TMA) units for asynchronous read/write acceleration of tensor data, further expediting neural network tasks. These technological transformations have garnered significant attention from numerous industrial and academic researchers.

AMD has also integrated matrix cores and FP16/BF16 data types into the architecture of its GPGPU series chips to support accelerated neural network tasks. Intel, in its upcoming Ponte Vecchio XMX chip architecture, has incorporated matrix engines and control flow processing prefetch units to support accelerated neural network tasks.

In traditional GPU architectures, NVIDIA has focused on accelerating applications such as neural graphics, metaverse, and digital humans. In its latest GeForce RTX 4090 chip, it collaboratively utilizes ray tracing cores (RTCore), tensor cores, and general-purpose computing cores to realize 3D scene enhancement based on deep learning and the DLSS technology introduced in Section \ref{5}. AMD has also incorporated matrix cores and ray tracing accelerators into its high-end GPU series chips to support relevant field applications. Additionally, the Special Function Unit (SFU) within the GPU can accelerate transcendental functions (such as reciprocal, square root, power functions, logarithms, trigonometric functions). Since the hardware acceleration of SFU essentially employs a method combining table lookup with double interpolation, it can also support hash encoding table lookup algorithms.
\subsubsection{Domain-Specific Architecture (DSA)}
With the continuous advancement of deep learning technologies, the demand for underlying hardware computational power in upper-layer application scenarios has experienced explosive growth. John \emph{et al.} \cite{b79} proposed Domain-Specific Architecture (DSA) as a response through collaborative software-hardware design. DSA is a type of architecture more inclined towards hardware-centric, custom-designed structures for specific problem domains. DSA can provide significant performance (and energy efficiency) gains for the domain, being more aligned with the application requirements compared to general-purpose processors. It achieves better performance when accelerating specific applications, while also possessing a certain level of programmability. Thus, compared to Application-Specific Integrated Circuits (ASIC), DSA offers greater flexibility and functional coverage.

Neural Processing Units (NPUs) or Tensor Processing Units (TPUs) are DSA architectures tailored for neural networks. Several commercial NPU/TPU hardware platforms have emerged. Reuther \emph{et al.} \cite{b88} have compiled and summarized commercially available neural network accelerators with peak performance and power consumption figures. Table. \MakeUppercase{\romannumeral 9} lists some NPU/TPU platforms and their support for neural network acceleration.
\begin{table*}[!ht]
    \centering
    \caption{\textbf{The Acceleration for NN in NPU/TPU}}
    \label{table9}
    \scalebox{1.08}{
    \begin{tabular}{|l|l|l|l|l|}
    \hline
        Platform & Supported Network Types & Data Types & Acceleration Structure & Scene-oriented \\ \hline
        TPUv4\cite{b89} & MLP & BF16 & Pulsed Array & Training/Inference  \\ 
        ~ & CNN & INT8 & ~ & ~ \\ 
        ~ & RNN & ~ & ~ & ~ \\ 
        ~ & Transformer & ~ & ~ & ~ \\ \hline
        TPUv4i\cite{b90} & MLP & BF16 & Pulsed Array & Inference  \\ 
        ~ & CNN & INT8 & ~ & ~ \\ 
        ~ & RNN & ~ & ~ & ~ \\ 
        ~ & Transformer & ~ & ~ & ~ \\ \hline
        Cambricon MLU370\cite{b91} & MLP & FP32 & Dot Product Tree & Training/Inference  \\ 
        ~ & CNN & FP16 & ~ & ~ \\ 
        ~ & RNN & BF16 & ~ & ~ \\ 
        ~ & Transformer & INT16/8/4 & ~ & ~ \\ \hline
        Baidu Kunlun Chip 2nd Generation \cite{b92} & CNN & FP16 & Multiply-Accumulate Array & Training/Inference  \\ 
        ~ & Transformer & INT8 & ~ & ~ \\ 
        ~ & GEMM & ~ & ~ & ~ \\ \hline
        ARM Ethos-U55\cite{b93} & CNN & INT8 & Multiply-Accumulate Unit & Inference  \\ 
        ~ & RNN & ~ & ~ & ~ \\ \hline
    \end{tabular}
    }
\end{table*}
NPU/TPU architectures are specifically designed for neural networks, primarily targeting deep neural networks and convolutional neural networks, supporting the limited operators and data types required for neural network computations. NPUs/TPUs designed for training typically exhibit higher computational capabilities and abundant memory resources, efficiently supporting tensor computation acceleration. NPUs/TPUs designed for inference, on the other hand, are usually constrained in terms of computational capability and memory bandwidth, with limited resources allocated for non-linear operations such as activation function processing, while other computational capabilities are relatively deficient.

The NePU, Neural Photonics Processing Unit, is a Domain-Specific Architecture (DSA) designed for tasks related to Neural Radiance Fields (NeRF) and is currently in the early stages of research. ICARUS \cite{b94} represents a specific architecture tailored for NeRF rendering. Fig. \ref{fig12} illustrates the overall architecture of ICARUS, which, together with the CPU and memory, forms the rendering system. The NeRF rendering process executed by ICARUS involves the following steps: 
\begin{itemize}
    \item storing input such as network models, encoding frequencies, and positional directions in memory, which is controlled by the CPU during runtime to provide input to ICARUS;
    \item distributing input data through on-chip buses to corresponding target plenoptic cores;
    \item loading network models, position, and directional data into ICARUS for NeRF processing (from the Position Encoding Unit (PEU) to the Multi-Layer Perceptron (MLP) engine and then to the Volumetric Rendering Unit (VRU)), where a group of rays (including all stepping points of the ray) is processed by the same plenoptic core;
    \item streaming out the final rendered pixel colors via the data bus.
\end{itemize}
The author enhances hardware acceleration efficiency by transforming the operations of the Fully Connected (FC) layer of the MLP computation into approximable Reconfigurable Multiple Constant Multiplications (RMCMs), thereby reducing approximately one-third of hardware complexity (including computational load and parameter capacity) compared to traditional multiply-accumulate computations. This enables the entire NeRF pipeline to be completed internally within the plenoptic core, with all weight parameters stored in the plenoptic core’s SRAM without the need for data exchange with DRAM. ICARUS, fabricated on a 40nm process node, occupies a chip area of 16.5 mm$^{2}$, achieves an operating frequency of 400 MHz, and requires 45.75 seconds to render an image of 800×800 resolution (with 192 sampling points for ray stepping), consuming 282.8mW. While the ICARUS architecture represents a certain level of innovation, there still exists a significant gap to achieve real-time rendering tasks for NeRF.

The Neural Graphics Processing Cluster (NGPC) \cite{b78} integrates a Neural Fields Processor (NFP) into the traditional GPU architecture to accelerate neural graphics applications. The overall architecture of NGPC is depicted in Fig. \ref{fig13}. The NGPC containing NFP is embedded as a hardware unit parallel to the Graphics Processing Cluster (GPC) within the GPU architecture to achieve on-chip heterogeneous computing. The NFP comprises an encoding engine (enc engine) and an MLP engine, which respectively accelerate input encoding and MLP computations. However, GPC units are required to handle hash encoding query computations. Through experimentation, the author determined that NGPC configurations containing 8, 16, and 32 NFP units deliver performance improvements of 12.94x, 20.85x, and 33.73x, respectively, compared to the RTX 3090 platform when executing NeRF, NSDF, GIA, and NVR applications. The author indicates that NGPC can achieve 30 FPS rendering at 4K ultra-high-definition resolution for NeRF applications (without explicitly specifying the quantity of NFP units). Estimations for the 7nm process node indicate that an NGPC-8 with a single NFP unit incurs approximately a 4.52\% increase in GPU area and a 2.75\% increase in power consumption. NGPC presents a method for achieving heterogeneous acceleration of neural rendering based on GPU architecture, leveraging the GPU’s existing general computing resources such as ray stepping and transcendental functions. However, this approach may lead to further hardware overhead and potentially exacerbate the occurrence of dark silicon phenomena in dedicated neural rendering scenarios.

\subsection{Design Challenges}
In summary, there are still certain design challenges for current hardware platforms when handling neural rendering applications:
\begin{itemize}
    \item Mainstream CPU platforms have been specifically designed and optimized for operations such as convolutions and matrix-matrix multiplications required by neural networks. However, for neural rendering tasks, CPU platforms are relatively constrained by neural network computational power and general-purpose computing capabilities, making them more suitable for neural forward rendering tasks or the inferencing portion of inverse rendering. Furthermore, due to the lack of rapid table lookup capabilities (activation functions, hash encoding), completing end-to-end high-definition real-time neural inverse rendering tasks poses significant challenges for CPU platforms.
    \item Currently, leading manufacturers such as AMD and NVIDIA have bifurcated their GPU chips into two distinct branches: GPGPU and traditional GPU platforms, catering separately to neural network/high-performance applications and traditional graphics applications. GPGPU platforms typically possess extremely high neural network computational power, high memory bandwidth, and on-chip storage capacity. However, they have omitted units such as rasterization, rendering output, and ray tracing found in traditional GPUs. GPGPU platforms can effectively accelerate neural network and volume rendering computations, well suiting the training requirements of neural rendering tasks. Nonetheless, they slightly lack specialized computational support for ray marching, hash code table lookup, and similar tasks. On the other hand, traditional GPU platforms, built upon traditional rasterization and ray tracing units, have augmented neural network computational power and higher memory performance. They can adapt well to the demands of neural forward rendering tasks but are unable to meet the acceleration requirements of neural inverse rendering tasks. Additionally, the development paths of these two product branches have increased manufacturers' research and development costs and system complexity.
    \item For DSA platforms, NPUs/TPUs are specific architectures designed for neural networks, primarily targeting deep neural networks and convolutional neural networks. Most hardware supports only the operations and data types required for neural network computations, with on-chip computational resources primarily dedicated to accelerating tensor computations, with a small portion allocated for non-linear operations such as activation function processing. Their general computing capabilities are relatively limited. Therefore, for neural rendering applications, due to the lack of computational power for ray marching, hash encoding, volume rendering computations, and the limitations in supported operations and data types, NPUs/TPUs hardware platforms may only be able to support a small number of neural rendering tasks. Research on various NePU architectures is mainly focused on NeRF-like neural rendering tasks and has achieved a certain degree of architectural breakthrough. However, they are commonly implemented as supplementary acceleration platforms for CPUs and GPUs, with certain limitations in specific scenarios. For instance, ICARUS \cite{b94} and the Instant 3D accelerator \cite{b67} require the use of a host SoC, while NGPC \cite{b78} adds complexity to GPU internal task scheduling and data flow mapping.
\end{itemize}

\section{Research and Development Trends}
This section provides a prospective analysis of the research and development directions for neural rendering systems from the perspectives of neural rendering applications and hardware acceleration architectures.
\subsection{Neural Rendering Applications}
After several years of development, neural rendering has demonstrated remarkable capabilities in scene representation, real-time global illumination, novel view synthesis, relighting, and spatial illumination. However, challenges such as generalization, scalability, and multimodal synthesis persist. Neural rendering applications have played a pivotal role in the advancement of Augmented Reality (AR) and Virtual Reality (VR). It is believed that neural rendering can entirely replace various submodules in traditional graphics rendering pipelines, such as surface subdivision or rasterization, enabling end-to-end neural forward rendering pipelines and the development of graphics applications based on neural rendering. In inverse rendering applications, current neural rendering works primarily focus on simple objects and relatively uncomplicated composite scenes. Challenges remain in extending methods developed based on single objects to large-scale scenes \cite{b22}, complex environmental scenarios \cite{b95}, or dynamic scenes \cite{b96}. Additionally, new view synthesis tasks still rely on large-scale multi-view datasets or are limited to training for specific objects, highlighting the importance of enhancing the generality of neural voxel representations across scenes or with a limited number of views. Multimodal learning applications of neural rendering imply simultaneous processing of semantic, textual, auditory, and visual signals. Rendering dynamic interactions and voice-face matching for digital humans pose significant challenges. Meeting consumer demands for neural rendering involves delineating functionalities and interaction modes for cloud-based and mobile-based applications, achieving lightweight and cost-effective deployment.
\subsection{Neural Rendering Processors}
In light of the high parallelism and specificity of neural rendering tasks, as well as the differences between neural rendering tasks and traditional graphics tasks, and the current hardware platform's limitations in handling neural rendering tasks, the research and development of Neural Rendering Processors (NRPU) to accelerate neural rendering applications represent a crucial developmental trend.

\subsubsection{Collaborative Design}
The collaborative design of software and hardware stands as a core approach in architectural design. It involves simplifying hardware implementation by partitioning system tasks into software and hardware functionalities, designing hardware acceleration units to address software algorithm bottlenecks, and seeking a balance between the two—a pursuit of paramount importance for designers. Neural network systems, exemplified by TPU, have provided us with much inspiration and experiential lessons in collaborative software and hardware design. For the future collaborative design of neural rendering systems, attention must be paid to the following issues:

1) The design objectives encompass neural rendering applications, a full-stack toolchain based on neural rendering pipelines for graphic editing, and neural rendering processors. This aims to innovate existing graphics processors and graphic editing tools, effectively supporting future key development areas such as virtual reality, augmented reality, film and television production, digital entertainment, artificial intelligence, and the metaverse.
2) Optimization methods should be determined by deployment scenarios. For instance, both Instant 3D \cite{b67} and GEN-NeRF \cite{b68} represent collaborative designs for NeRF-type applications based on AR/VR devices. As AR/VR devices operate within computational, memory, and power constraints, their collaborative design objectives focus on reducing the computational or storage overhead of algorithms and hardware while striving to maintain algorithm accuracy within an acceptable range or, if possible, lowering accuracy under such constraints.
3) Reasonable division of sub-functionality between software and hardware is essential. For instance, in NeRF-type neural rendering tasks, hardware acceleration is employed for sub-functions such as hash table computation, MLP computation, and volume rendering computation, while in RenderNet-type neural rendering tasks, hardware acceleration is utilized for sub-functions such as neural network computation or projection computation. On the other hand, software handles sub-functions like coordinate preprocessing, positional encoding, and ray-sampling point determination.
4) Hardware architecture design should fully consider algorithm maturity and hardware design cycles, retaining a certain degree of design margin and flexibility. Jouppi et al. \cite{b90} proposed ten crucial experiential lessons in TPU design, highlighting the importance of providing dynamic margins for DSAs to remain effective throughout their entire lifecycle, as well as the significance of flexibility in DSA design, as evidenced by the rapid replacement of a large number of LSTMs by Transformer-type applications.
\subsubsection{Scene Segmentation}
In anticipation of potential deployment scenarios for future neural rendering systems, the design of neural rendering processors is discussed separately for cloud and device scenarios:
\begin{itemize}
    \item Integrated Training and Rendering NRPU is tailored for cloud scenarios, deployed as neural rendering processors (accelerator cards) in data centers. In this setup, neural rendering applications accomplish neural rendering network training and rendering in the cloud or handle the re-computational workload during rendering. This processor can independently execute rendering tasks or transmit streams or specific encoded files over the network to be completed for final rendering on end-device platforms.
    \item Rendering-Specific NRPU is aimed at device scenarios, deployed within mobile devices or AR/VR devices as neural rendering processors (system-on-chip). In this context, neural rendering applications carry out image rendering at the device end. The rendering-specific NRPU may receive streams from the cloud or specific encoded files, and, through dedicated hardware acceleration, synthesize images and render output.
\end{itemize} 

\subsubsection{Architecture Design}
The integrated NRPU for rendering and training needs to possess powerful neural network computing capabilities and memory access performance similar to GPGPU/high-end NPU, effectively supporting the training of neural rendering networks. Simultaneously, it also requires robust specialized neural rendering computational capabilities for operations such as transcendental function processing, hash encoding table processing, and ray marching processing. Designing a programming model and architecture to meet the training and rendering data flow of neural rendering applications, as well as increasing the parallelism of the hardware acceleration units for neural rendering while designing performance-matched memory access and data transmission paths and on-chip storage space, will be a significant challenge within the constraints of transistor capacity and process technology.

Rendering-specific NRPU needs to determine the operator units, peak computing performance, storage capacity, and bandwidth required by the hardware based on specific neural rendering algorithms. Additionally, it necessitates data flow and performance analysis of competition for memory access bandwidth and network transmission bandwidth among NRPU and other on-chip system hardware units (such as CPU, ISP, encoding/decoding units, etc.). Efficiently mapping the neural rendering pipeline to the on-chip system to enhance on-chip data reuse and reduce memory access bandwidth requirements, as well as reducing hardware operating power consumption and design costs while ensuring high-quality real-time neural rendering, presents a significant challenge in device-end platforms constrained by power consumption, area, and price.
\section{Conclusion}
With the rapid rise and development of deep neural networks, neural rendering, as a fusion method of deep learning and computer graphics, has emerged. Neural rendering is an image and video generation method based on deep learning, combining deep neural network models with the physical knowledge of computer graphics to obtain controllable and realistic scene models, enabling control of scene attributes such as lighting, camera parameters, and posture. In recent years, neural rendering has made rapid progress, with applications ranging from enhancing forward rendering effects using deep neural networks to new view synthesis, shape and material editing, relighting, and avatar generation in inverse rendering. This paper introduces representative research achievements of neural rendering in forward rendering, inverse rendering, and post-processing applications, analyzes in detail the common hardware acceleration requirements of neural rendering applications, and finally discusses the design challenges of neural rendering processor architecture according to scenarios.

\bibliographystyle{ieee}
\bibliography{access.bib}

\EOD

\end{document}